\documentclass[5p,sort&compress]{elsarticle}
\usepackage{epstopdf}
\usepackage{cuted}
\usepackage{subfigure,graphicx}
\usepackage{float}
\usepackage{amsmath, amsfonts, amssymb}
\usepackage{amsthm}
\usepackage{epsf}
\usepackage{amsfonts}
\usepackage{stmaryrd}
\usepackage{amssymb}
\usepackage{leftidx}
\usepackage{color}
\usepackage{mathtools}
\usepackage{placeins}
\usepackage{booktabs}
\usepackage{enumitem}
\usepackage{caption}
\usepackage{tabu}
\usepackage{multirow}
\usepackage{stackengine}
\usepackage[utf8]{inputenc}
\usepackage[english]{babel}
\usepackage{color}

\theoremstyle{plain}
\newtheorem{theorem}{Theorem}

\newtheorem{remark}[theorem]{Remark}

\def\bfx{{\bf x}}
\def\bfy{{\bf y}}

\def\bfC{{\bf C}}

\def\bfI{{\bf I}}

\def\bfN{{\bf N}}

\def\bfS{{\bf S}}
\def\bfT{{\bf T}}
\def\bfX{{\bf X}}
\def\bfF{{\bf F}}

\def\bfD{{\bf D}}
\def\bfe{{\bf e}}

\def\Ktan{\mbox{\boldmath$\mathcal{K}$}}

\def\e0{\varepsilon_0}

\def\s0{\sigma_0}

\long\def\symbolfootnote[#1]#2{\begingroup%
\def\thefootnote{\fnsymbol{footnote}}\footnote[#1]{#2}\endgroup}

\begin{document}
\begin{frontmatter}

\title{The ``pure-shear'' fracture test for viscoelastic elastomers and \\ its revelation on Griffith fracture\vspace{0.1cm}}

\author{Bhavesh Shrimali}
\ead{bshrima2@illinois.edu}

\author{Oscar Lopez-Pamies}
\ead{pamies@illinois.edu}

\address{Department of Civil and Environmental Engineering, University of Illinois, Urbana--Champaign, IL 61801, USA  \vspace{0.2cm}}

\vspace{0.2cm}

\begin{abstract}

\vspace{0.2cm}

Strikingly, ``pure-shear’’ fracture tests have repeatedly shown that fracture nucleation in (common hydrocarbon and other types of) viscoelastic elastomers occurs at a critical stretch that is independent of the stretch rate at which the test is carried out. In this Letter, we demonstrate that this remarkable --- yet overlooked --- experimental finding implies that the Griffith criticality condition that governs nucleation of fracture from large pre-existing cracks in viscoelastic elastomers can be written in fact as an expression \emph{not} in terms of an elusive loading-history-dependent critical tearing energy $T_c$, as ordinarily done, but as one exclusively in terms of the intrinsic fracture energy $G_c$ of the elastomer.

\vspace{0.2cm}

\endkeyword

\end{abstract}

\end{frontmatter}

\section{Introduction and main result}\label{Sec:IntroMain}

Following in the footstep of Griffith \cite{Griffith21} and Busse \cite{Busse34}, Rivlin and Thomas \cite{Rivlin1953} famously identified three types of tests --- the so-called ``pure-shear'', single-edge-crack-tensile, and ``trousers'' tests --- that allowed for an expedient study of when fracture nucleates from large pre-existing cracks in elastomers subjected to quasi-static deformations. While Rivlin and Thomas' analysis of these tests made critical use of the idealization that elastomers are purely elastic solids, the flurry of subsequent investigations that their pioneering work triggered extended the same approach to account for the fact that elastomers are inherently viscoelastic solids
\cite{Greensmith55,Mullins59,Thomas00,Knauss15,LakeandThomas67}. In so doing, they established that fracture may nucleate from a large pre-existing crack in an elastomer whenever the change in \emph{total} deformation (stored and dissipated) energy $\mathcal{W}$ in the bulk with respect to an added surface area to the pre-existing crack $\Gamma_0$ reaches a certain critical tearing energy $T_c$ characteristic of the elastomer:
\begin{equation}
-\dfrac{\partial \mathcal{W}}{\partial \Gamma_0}=T_{c}.\label{Tc-0}
\end{equation}
In this expression, the added surface area refers to the undeformed configuration and the derivative is to be carried out under fixed boundary conditions on the parts of the boundary which are not traction-free. Notably, $T_c$ is \emph{not} a constant but --- much like $\mathcal{W}$ --- a function of the loading history. Physically, $T_c$ describes the \emph{total} energy (per unit fracture area) expended in the tearing process and thus it contains contributions from the actual creation of new surface as well as from the viscous dissipation (assuming that there are no other dissipation mechanisms, such as strain crystallization) taking place around the crack front and the rest of the bulk.

Experiments carried out at extremely low loading rates, at high temperatures, and on solvent-swollen specimens, when viscous effects are minimized, have shown that
\begin{equation*}
T_{c}=G_c,
\end{equation*}
where $G_c$ denotes the intrinsic fracture energy, or critical energy release rate, associated with the creation of new surface in the given elastomer \cite{LakeandThomas67,Knauss71,Ahagon-Gent75,Gent82,GentLai94}. It is a material constant, independent of time. Its value is in the same range
\begin{equation}
G_c\in[10,100]\, {\rm N}/{\rm m}\label{Gc-range}
\end{equation}
for all common hydrocarbon elastomers \cite{Ahagon-Gent75,Gent82}.

More generally, experiments carried out at various loading rates, when viscous effects are not negligible and could even be dominant, have shown that
\begin{equation*}
T_{c}=G_c(1+f_c),
\end{equation*}
where $f_c$ is a non-negative function of the loading history that scales with the viscosity of the elastomer at hand \cite{Mullins59,GentLai94,Knauss1973,Gent96}. {\color{black} Precisely how $f_c$ --- and hence $T_c$ --- depends on the loading history has remained an open problem for decades, save for the few specific loading conditions (such as deformations applied at constant stretch rates in ``pure-shear'' fracture tests) that have allowed to directly measure $T_c$ experimentally. This lacuna in knowledge has severely hindered the practical utility of the Griffith criticality condition (\ref{Tc-0}).}

In this Letter, we show that the Griffith criticality condition (\ref{Tc-0}) can be reduced in fact to a more fundamental and useful form that involves \emph{not} the elusive critical tearing energy $T_c$, but only the intrinsic fracture energy $G_c$ of the elastomer. We do so by combining two elementary observations:

\begin{enumerate}[label=(\roman*)]

\item{For a viscoelastic elastomer, without loss of generality, the total deformation energy $\mathcal{W}$ in (\ref{Tc-0}) can be written in the form\footnote{Rheological representations are helpful to make this partition of energies apparent. For instance, in the rheological representation depicted in Fig. \ref{Fig2}, $\mathcal{W}^{{\rm Eq}}$ and $\mathcal{W}^{{\rm NEq}}$ correspond to the elastic energy stored in the equilibrium and non-equilibrium springs, whereas $\mathcal{W}^{v}$ corresponds to the viscous energy dissipated by the dashpot.}
\begin{equation}
\mathcal{W}=\underbrace{\mathcal{W}^{{\rm Eq}}+\mathcal{W}^{{\rm NEq}}}_\text{stored}+\underbrace{\mathcal{W}^{v}}_\text{dissipated} \label{WWW}
\end{equation}
as a sum of a stored part and a dissipated part. The stored part of the energy is comprised itself of two parts: an equilibrium part $\mathcal{W}^{{\rm Eq}}$ and a non-equilibrium part $\mathcal{W}^{{\rm NEq}}$. The latter represents the part of the stored energy that gets dissipated via viscous deformation eventually. On the other hand, $\mathcal{W}^{v}$ represents the part of the energy that is dissipated via viscous dissipation instantaneously. Granted (\ref{WWW}), the criticality condition (\ref{Tc-0}) can be rewritten as
\begin{equation}
-\dfrac{\partial \mathcal{W}^{{\rm Eq}}}{\partial \Gamma_0}=G_c+G_c \,f_c+\dfrac{\partial \mathcal{W}^{{\rm NEq}}}{\partial \Gamma_0}+\dfrac{\partial \mathcal{W}^{v}}{\partial \Gamma_0}. \label{Tc-1}
\end{equation}
In view of this relation, upon noticing the string of inequalities $G_{c}\, f_{c}\geq 0$, $\partial \mathcal{W}^{{\rm NEq}}/\partial \Gamma_0\leq 0$, $\partial \mathcal{W}^{v}/\partial \Gamma_0$ $\leq 0$, and that, much like $f_c$, the terms $\partial \mathcal{W}^{{\rm NEq}}/\partial \Gamma_0$ and $\partial \mathcal{W}^{v}/\partial \Gamma_0$ scale with the viscosity of the elastomer, one may naturally wonder whether
\begin{equation}
G_c \,f_c+\dfrac{\partial \mathcal{W}^{{\rm NEq}}}{\partial \Gamma_0}+\dfrac{\partial \mathcal{W}^{v}}{\partial \Gamma_0}= 0 \label{G-visc}
\end{equation}
and hence whether the criticality condition (\ref{Tc-0}) is, in point of fact, given by
\begin{equation*}
-\dfrac{\partial \mathcal{W}^{{\rm Eq}}}{\partial \Gamma_0}=G_{c}.\vspace{-0.7cm}
\end{equation*}
%
%
\begin{figure}[h!]
   \centering \includegraphics[width=2.4in]{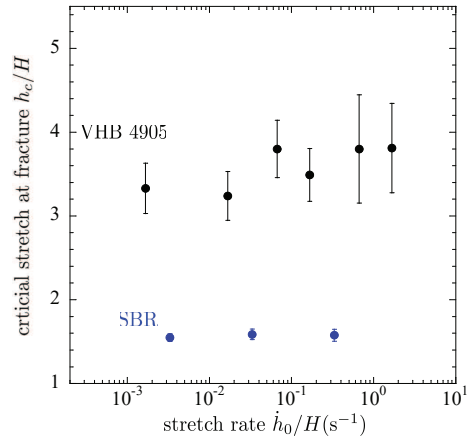}
   \caption{Critical stretch $h_c/H$, as a function of the applied stretch rate $\dot{h}_0/H$, at which fracture occurs in the ``pure-shear'' tests reported in \cite{Cai17} for a SBR rubber, a common hydrocarbon elastomer, and in \cite{Pharretal2012} for VHB 4905, an acrylic elastomer.}\label{Fig0}
\end{figure}
%
}

\item{Experiments have repeatedly shown that fracture in ``pure-shear'' tests of viscoelastic elastomers occurs at a critical stretch that is independent (to within experimental error) of the stretch rate at which the test is carried out. This appears to be the behavior of common hydrocarbon elastomers \cite{Major10,Cai17}, as well as that of more modern types of elastomers \cite{Pharretal2012,Kangetal2020}.

    As examples of illustrative experimental data, Fig. \ref{Fig0} reproduces the results reported in \cite{Cai17} for a SBR rubber, a common hydrocarbon elastomer, and in \cite{Pharretal2012} for VHB 4905, an acrylic elastomer.
}

\end{enumerate}

\paragraph{The main result} As elaborated in the next two sections, the remarkable experimental fact that the critical stretch at which fracture occurs in ``pure-shear'' tests is independent of the applied stretch rate necessarily implies that relation (\ref{G-visc}) is indeed correct and hence that the criticality condition
\begin{equation}
-\dfrac{\partial \mathcal{W}^{{\rm Eq}}}{\partial \Gamma_0}=G_{c}\label{Gc-0}
\end{equation}
is the {\color{black} fundamental form of the} Griffith condition that governs the nucleation of fracture from large pre-existing cracks in viscoelastic elastomers.

\section{Global analysis of the ``pure-shear'' test}\label{Sec:Global}

As already alluded to above, Rivlin and Thomas \cite{Rivlin1953} famously identified the ``pure-shear'' test as one of the most convenient tests to study nucleation of fracture from large pre-existing cracks in elastomers, this provided that elastomers are viewed as purely elastic solids. As will become apparent in this section, Rivlin and Thomas' analysis of the ``pure-shear'' test can be easily transcribed to viscoelastic elastomers.

%
\begin{figure}[t!]
   \centering \includegraphics[width=3.3in]{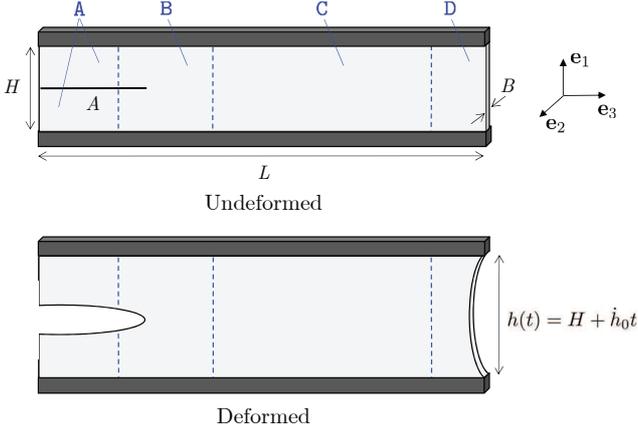}
   \caption{Schematic of the ``pure-shear'' test for a viscoelastic elastomer carried out at constant global stretch rate $\dot{h}_0/H$. The dimensions in the undeformed configuration are such that $B\ll H\ll A \ll L$. The region $\texttt{C}$ in the specimen is essentially in a state of spatially uniform pure shear, hence the name of the test.}\label{Fig1}
\end{figure}
%
Consider the ``pure-shear'' test schematically depicted in Fig. \ref{Fig1}, where the thickness of the specimen --- taken to be made of an isotropic incompressible viscoelastic elastomer --- is much smaller than its height ($B\ll H$), its height is much smaller than its length ($H\ll L$), and the initial length of the pre-existing crack is much larger than the height of the specimen but much smaller than its length ($H\ll A\ll L$). In other words, the specimen is essentially an infinitely long strip that contains a semi-infinitely long edge crack.

A load is applied by separating the top and bottom grips at a constant rate $\dot{h}_0$ so that, as a function of time $t\in[0,T]$, the current height of the specimen is given by the relation $h(t)=H+\dot{h}_0 t$. It follows that at any given time $t\in(0,T]$, because of the special geometry of the specimen and the incompressibility of the elastomer, the specimen features four different regions of deformation; see Fig. \ref{Fig1}. Adhering to the original region labeling used by Rivlin and Thomas (see Fig. 8 in \cite{Rivlin1953}), the region $\texttt{A}$ is substantially undeformed, the crack-front region $\texttt{B}$ and the fringe region $\texttt{D}$ are in a complex state of deformation (highly non-uniform in space), while region $\texttt{C}$ is in a state of spatially uniform pure shear.

Now, for a fixed loading rate $\dot{h}_0$ and a fixed time $t$, so that the separation between the grips is fixed at $h(t)=H+\dot{h}_0 t$, consider an increase in the crack surface of amount ${\rm d}\Gamma_0=B {\rm d}A$. This increase in crack surface does \emph{not} alter the complex state of deformation in $\texttt{B}$ but, instead, simply shifts this entire region in the direction of the added crack, resulting in the growth of region $\texttt{A}$ at the expense of region $\texttt{C}$. In other words, an added crack ${\rm d}\Gamma_0$ results in the transferring of a volume $H{\rm d}\Gamma_0$ of the specimen from a state of pure shear to the undeformed state. Making use of this observation, we have that the change in total (stored and dissipated) deformation energy in the bulk with respect to the added crack in a ``pure-shear'' test can be readily computed in terms of the spatially uniform pure-shear behavior of the elastomer in region $\texttt{C}$, precisely,
\begin{equation}
-\dfrac{\partial \mathcal{W}}{\partial \Gamma_0}=H\displaystyle\int_{1}^{\frac{h(t)}{H}}S_{ps}(\lambda;\dot{\lambda}_{0})\,{\rm d}\lambda,\label{PS-WG0}
\end{equation}
where $S_{ps}(\lambda;\dot{\lambda}_0)$ denotes the stress-stretch relation of the given elastomer under pure shear applied at the constant stretch rate $\dot{\lambda}_0=\dot{h}_0/H$, that is, under a spatially uniform deformation gradient of the form $\bfF={\rm diag}(\lambda,\lambda^{-1},1)$, with $\lambda=1+\dot{\lambda}_{0} t$, and first Piola-Kirchhoff stress tensor of the form $\bfS={\rm diag}(S_{ps},0,S_{lat})$ with respect to the laboratory frame of referenced indicated in Fig. \ref{Fig1}.

For any viscoelastic elastomer, it so happens that we can write the stress-stretch relation in the additive form
\begin{equation*}
S_{ps}(\lambda;\dot{\lambda}_{0})=S^{{\rm Eq}}_{ps}(\lambda)+S^{{\rm NEq},v}_{ps}(\lambda;\dot{\lambda}_{0}),
\end{equation*}
where $S^{{\rm Eq}}_{ps}$ stands for the stress associated with the equilibrium part of the underlying elastic energy, while $S^{{\rm Eq},v}_{ps}$ denotes the stress associated with the non-equilibrium part of the elastic energy and the dissipated viscous energy. By making use of this decomposition, relation (\ref{PS-WG0}) can then be rewritten as
\begin{equation*}
-\dfrac{\partial \mathcal{W}}{\partial \Gamma_0}=H\displaystyle\int_{1}^{\frac{h(t)}{H}}S^{{\rm Eq}}_{ps}(\lambda)\,{\rm d}\lambda+H\displaystyle\int_{1}^{\frac{h(t)}{H}}S^{{\rm NEq}, v}_{ps}(\lambda;\dot{\lambda}_{0})\,{\rm d}\lambda.
\end{equation*}
Direct use of this last result in the general criticality condition (\ref{Tc-1}) leads to
\begin{equation}
\underbrace{H\displaystyle\int_{1}^{\frac{h(t)}{H}}S^{{\rm Eq}}_{ps}(\lambda)\,{\rm d}\lambda}_{-\dfrac{\partial \mathcal{W}^{{\rm Eq}}}{\partial \Gamma_0}}=G_c+G_c \,f_c-\underbrace{H\displaystyle\int_{1}^{\frac{h(t)}{H}}S^{{\rm NEq}, v}_{ps}(\lambda;\dot{\lambda}_{0})\,{\rm d}\lambda}_{-\left(\dfrac{\partial \mathcal{W}^{{\rm NEq}}}{\partial \Gamma_0}+\dfrac{\partial \mathcal{W}^{v}}{\partial \Gamma_0}\right)}. \label{Tc-PS}
\end{equation}

At this point, we can make two critical observations. By virtue of the independence of $S^{{\rm Eq}}_{ps}(\lambda)$ of $\dot{\lambda}_0$, the change in equilibrium elastic energy $-\partial \mathcal{W}^{{\rm Eq}}/\partial \Gamma_0$ in (\ref{Tc-PS}) --- much like the material constant $G_c$ --- is independent of the stretch rate $\dot{\lambda}_0$. By contrast, the change in non-equilibrium elastic energy $-\partial \mathcal{W}^{{\rm NEq}}/\partial \Gamma_0$ and dissipated viscous energy $-\partial \mathcal{W}^{v}/\partial \Gamma_0$ --- much like the term $G_c f_c$ --- \emph{do} depend on the stretch rate $\dot{\lambda}_0$. These behaviors, when combined with the experimental fact that viscoelastic elastomers in ``pure-shear'' tests carried out at constant stretch rates nucleate fracture at the same global stretch $h(t)/H$, necessarily imply that relations (\ref{G-visc}) and (\ref{Gc-0}) must hold true, for the equality in (\ref{Tc-PS}) can be satisfied at fixed $h(t)/H$ for all stretch rates $\dot{\lambda}_0$ only when the stretch-rate-dependent part
\begin{equation*}
G_c \,f_c-H\displaystyle\int_{1}^{\frac{h(t)}{H}}S^{{\rm NEq}, v}_{ps}(\lambda;\dot{\lambda}_{0})\,{\rm d}\lambda=0
\end{equation*}
and the stretch-rate-independent part
\begin{equation*}
H\displaystyle\int_{1}^{\frac{h(t)}{H}}S^{{\rm Eq}}_{ps}(\lambda)\,{\rm d}\lambda-G_c=0.
\end{equation*}

\section{Full-field analysis of the ``pure-shear'' test}\label{Sec:Full-field}

Complementary to the global analysis presented above, in this section we present the full-field analysis of the ``pure-shear'' test for an isotropic incompressible elastomer with \emph{Gaussian elasticity} and \emph{constant viscosity}, which, arguably, is the most basic type of viscoelastic elastomer and thus can be viewed as a canonical problem. We begin by formulating the pertinent initial-boundary-value problem and then proceed with the presentation and discussion of the results.

\subsection{Formulation of the initial-boundary-value problem}

\subsubsection{Initial configuration and kinematics}

Consider rectangular specimens of length $L=152$ mm and height $H=10$ mm in the $\bfe_3$ and $\bfe_1$ directions and constant thickness $B=0.5$ mm in the $\bfe_2$ direction; see Fig. \ref{Fig1}. The specimens contain a pre-existing edge crack of five different lengths
\begin{equation*}
A=15,20,25,30,40\; {\rm mm}
\end{equation*}
in the $\bfe_3$ direction. These specific values for $L, H, B, A$ are chosen because they are representative of those typically used in experiments; see, in particular, \cite{Pharretal2012}. Here, $\{\bfe_i\}$ stands for the laboratory frame of reference. We place its origin at the specimens' midplane along the edge containing the crack so that, in their initial configuration at time $t=0$, the specimens occupy the domain
\begin{equation*}
\overline{\Omega}_0=\{\bfX: \bfX\in\mathcal{P}_0\setminus\Gamma_0\},
\end{equation*}
where
\begin{equation*}
\mathcal{P}_0=\left\{\bfX: |X_1|\leq\dfrac{H}{2},\,|X_2|\leq\dfrac{B}{2},\,  0\leq X_3\leq L  \right\}
\end{equation*}
and
\begin{equation*}
\Gamma_0=\left\{\bfX: X_1=0,\,|X_2|\leq\dfrac{B}{2},\,  0\leq X_3\leq A  \right\}.
\end{equation*}

In response to the applied boundary conditions described below, the position vector $\bfX$ of a material point in the specimens will move to a new position specified by
\begin{equation*}
\bfx=\bfy(\bfX, t),
\end{equation*}
where $\bfy$ is an invertible mapping from $\Omega_0$ to the current configuration $\Omega(t)$. Making use of standard notation, we write the deformation gradient and Lagrangian velocity fields at $\bfX$ and $t$ as
\begin{equation*}
\bfF(\bfX, t)=\nabla\bfy(\bfX,t)=\frac{\partial \bfy}{\partial \bfX}(\bfX,t)
\end{equation*}
and
\begin{equation*}
\textbf{V}(\bfX,t)=\dot{\bfy}(\bfX, t)= \frac{\partial \bfy}{\partial t}(\bfX,t);
\end{equation*}
the ``dot'' notation will be employed throughout to denote the Lagrangian time derivative (i.e., with $\bfX$ held fixed) of any field quantity.

\subsubsection{Constitutive behavior of the elastomer}

The specimens are taken to be made of a viscoelastic elastomer with Gaussian elasticity and constant viscosity. Precisely, making use of the two-potential formalism \cite{KLP16}, the constitutive behavior of the elastomer (for isothermal conditions) is characterized by the two thermodynamic potentials
\begin{equation}
\psi(\bfF,\bfF^v)=\left\{\begin{array}{ll}\underbrace{\dfrac{\mu}{2}\left[I_1-3\right]}_\text{$\psi^{{\rm Eq}}(\bfF)$}+\underbrace{\dfrac{\nu}{2}\left[I^e_1-3\right]}_\text{$\psi^{{\rm NEq}}\left(\bfF{\bfF^v}^{-1}\right)$}& {\rm if}\quad J=1\\ \\
+\infty & {\rm otherwise}\end{array}\right.\label{psi}
\end{equation}
and
\begin{equation}
\phi(\bfF,\bfF^v,\dot{\bfF}^v)=\left\{\begin{array}{ll}
\dfrac{1}{2}\dot{\bfF}^v{\bfF^v}^{-1}\cdot\left[2\,\eta\,\Ktan\,\dot{\bfF}^v{\bfF^v}^{-1}\right]& \vspace{0.2cm} \\
& \hspace{-1.5cm}{\rm if}\;{\rm tr}( \dot{\bfF}^v{\bfF^v}^{-1})=0 \\ \\
+\infty & \hspace{-0.2cm} {\rm otherwise}\end{array}\right., \label{phi}
\end{equation}
where $\psi$ and $\phi$ stand, respectively, for the free energy and dissipation potential describing how the elastomer stores and dissipates energy through elastic and viscous deformation. In these expressions, the second-order tensor $\bfF^v$ is an internal variable of state that describes roughly the ``viscous part'' of the deformation gradient $\bfF$,
\begin{align}
&I_1=\bfF\cdot\bfF={\rm tr}\,\bfC, \quad  J=\det\bfF=\sqrt{\det\bfC},\nonumber\\
& I_1^e=\bfF{\bfF^v}^{-1}\cdot\bfF{\bfF^v}^{-1}={\rm tr}(\bfC{\bfC^{v}}^{-1}),\label{invariants}
\end{align}
where $\bfC=\bfF^T\bfF$ denotes the right Cauchy-Green deformation tensor, $\bfC^v={\bfF^v}^T\bfF^v$, $\mathcal{K}_{ijkl}=\frac{1}{2}(\delta_{ik}\delta_{jl}+\delta_{il}\delta_{jk}-\frac{2}{3}\delta_{ij}\delta_{kl})$ stands for the standard deviatoric orthogonal projection tensor, and $\mu\ge0$, $\nu\ge0$, $\eta\geq0$ are material constants.

\begin{figure}[h!]\centering
 \includegraphics[width=2.5in]{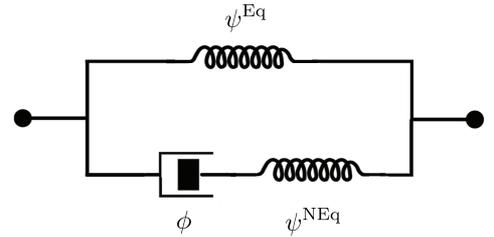}
\caption{\small Rheological representation of a viscoelastic elastomer.}
   \label{Fig2}
\end{figure}

For a complete account of the two-potential framework as it pertains to elastomers, the interested reader is referred to \cite{KLP16}. Here, it suffices to remark that the two-potential model (\ref{psi})-(\ref{phi}) corresponds to a generalization of the classical Zener or standard solid model \cite{Zener48} to the setting of  finite deformations. As schematically depicted by the rheological representation in Fig. \ref{Fig2}, the function $\psi^{{\rm Eq}}$ in (\ref{psi}) characterizes the Gaussian elastic energy storage in the elastomer at states of thermodynamic equilibrium, whereas $\psi^{{\rm NEq}}$ characterizes the additional Gaussian elastic energy storage at non-equilibrium states (that is, again, the part of the energy that gets dissipated eventually). On the other hand, the parameter $\eta$ in (\ref{phi}) characterizes the constant viscosity of the elastomer.

Granted the two thermodynamic potentials (\ref{psi}) and (\ref{phi}), it follows that the first Piola-Kirchhoff stress tensor $\bfS$ at any material point $\bfX\in\Omega_0$ and time $t\in[0,T]$ is expediently given by the relation \cite{KLP16}
\begin{equation}
\bfS(\bfX,t)=\frac{\partial \psi}{\partial\bfF}(\bfF,\bfF^v),\label{S-gen}
\end{equation}
where $\bfF^v$ is implicitly defined by the evolution equation
\begin{equation}
\dfrac{\partial \psi}{\partial \bfF^v}(\bfF,\bfF^v)+\dfrac{\partial \phi}{\partial \dot{\bfF}^v}(\bfF,\bfF^v,\dot{\bfF}^v)={\bf0}.\label{Evolution-gen}
\end{equation}
Making use of the specific forms (\ref{psi}) and (\ref{phi}), this relation can be rewritten more explicitly as
\begin{equation}
\bfS(\bfX,t)=\mu \bfF+\nu\bfF{\bfC^v}^{-1}-p\bfF^{-T},\label{S-Neo}
\end{equation}
where $p$ stands for the arbitrary hydrostatic pressure associated with the incompressibility constraint $J=1$ of the elastomer and $\bfC^v$ is defined implicitly as the solution of the evolution equation
\begin{equation}
\dot{\bfC}^v(\bfX,t)=\dfrac{\nu}{\eta}\left[\bfC-\dfrac{1}{3}\left(\bfC\cdot{\bfC^v}^{-1}\right)\bfC^v\right].  \label{Evolution-Neo}
\end{equation}
Note that the dependence on the internal variable $\bfF^v$ ends up entering (\ref{S-Neo}) and (\ref{Evolution-Neo}) only through the symmetric combination $\bfC^v={\bfF^v}^T\bfF^v$.

\begin{remark}\label{Remark1} The solid and fluid limiting cases of the constitutive behavior (\ref{S-Neo})-(\ref{Evolution-Neo}). {\rm The prototypical constitutive behavior (\ref{S-Neo})-(\ref{Evolution-Neo}) contains two important limiting cases. The first one, which corresponds to setting the elastomer viscosity either to $\eta=0$ or $\eta\rightarrow+\infty$, is that of a \emph{Gaussian elastic or Neo-Hookean solid}. The second one, which corresponds to setting the equilibrium and non-equilibrium moduli to $\mu=0$ and $\nu\rightarrow+\infty$, is that of a \emph{Newtonian fluid}.

To see the specialization to the elastic solid limiting case, note that when $\eta=0$, the solution to the evolution equation (\ref{Evolution-Neo}) is simply $\bfC^v=\bfC$ and the first Piola-Kirchhoff stress tensor (\ref{S-Neo}) reduces, with a slight abuse of notation, to $\bfS(\bfX,t)=\mu\bfF-p\bfF^{-T}$. Similarly, when $\eta\rightarrow+\infty$, the solution to the evolution equation (\ref{Evolution-Neo}) is $\bfC^v=\bfI+O(\eta^{-1})$ and the first Piola-Kirchhoff stress tensor (\ref{S-Neo}) reduces to $\bfS(\bfX,t)=(\mu+\nu)\bfF-p\bfF^{-T}$.

On the other hand, to see the specialization to the viscous fluid limiting case, note that when $\mu=0$ and $\nu\rightarrow+\infty$, the solution to the evolution equation (\ref{Evolution-Neo}) is given by $\bfC^v=\bfC+\nu^{-1}(-\eta\dot{\bfC}+p_1\bfC)+O(\nu^{-2})$ and the first Piola-Kirchhoff stress tensor (\ref{S-Neo}) reduces to $\bfS(\bfX,t)=\eta(\dot{\bfF}\bfF^{-1}\bfF^{-T}+\bfF^{-T}\dot{\bfF}^T\bfF^{-T})-q\bfF^{-T}$; in these last two expressions, $p_1$ and $q$ are arbitrary hydrostatic pressures associated with the incompressibility constraint. Accordingly, the Cauchy stress tensor $\bfT=\bfS\bfF^{T}$ specializes to $\bfT(\bfx,t)=2\eta\bfD-q\bfI$, where $\bfD=1/2(\dot{\bfF}\bfF^{-1}+\bfF^{-T}\dot{\bfF}^T)$ is the rate of deformation tensor.
}

\end{remark}

\subsubsection{Initial and boundary conditions}

In their initial configuration, we consider that the specimens are undeformed and stress-free. Therefore, we have the initial conditions
\begin{equation}
\left\{\begin{array}{l}
\bfy(\bfX,0)=\bfX\vspace{0.2cm}\\
p(\bfX,0)=\mu+\nu\vspace{0.2cm}\\
\bfC^v(\bfX,0)=\bfI \end{array}\right., \quad\bfX\in \overline{\Omega}_0.\label{ICs}
\end{equation}

Save for the top boundary
\begin{equation*}
\partial\Omega^{\mathcal{T}}_0=\left\{\bfX: X_1=\dfrac{H}{2},\,|X_2|\leq\dfrac{B}{2},\,  0\leq X_3\leq L  \right\}
\end{equation*}
and the bottom boundary
\begin{equation*}
\partial\Omega^{\mathcal{B}}_0=\left\{\bfX: X_1=-\dfrac{H}{2},\,|X_2|\leq\dfrac{B}{2},\,  0\leq X_3\leq L  \right\},
\end{equation*}
the entire boundary $\partial\Omega_0$ of the specimens is traction free. The top and bottom boundaries are separated in the $\bfe_1$ direction at the constant rate $\dot{h}_0$ so that, as a function of time $t\in[0,T]$, the current height of the specimen is given by the relation $h(t)=H+\dot{h}_0 t$. Precisely, making use of the notation $\textbf{s}(\bfX,t)=\bfS\bfN$, we have that
\begin{equation}
\left\{\hspace{-0.15cm}\begin{array}{ll}
y_1(\bfX,t)=X_1+\dfrac{\dot{h}_0}{2} t, & (\bfX,t)\in\partial\Omega^{\mathcal{T}}_0\times[0,T] \vspace{0.15cm}\\
y_3(\bfX,t)=X_3, & (\bfX,t)\in\partial\Omega^{\mathcal{T}}_0\times[0,T] \vspace{0.15cm}\\
s_2(\bfX,t)=0, & (\bfX,t)\in\partial\Omega^{\mathcal{T}}_0\times[0,T] \vspace{0.15cm}\\
y_1(\bfX,t)=X_1-\dfrac{\dot{h}_0}{2} t, & (\bfX,t)\in\partial\Omega^{\mathcal{B}}_0\times[0,T] \vspace{0.15cm}\\
y_3(\bfX,t)=X_3, & (\bfX,t)\in\partial\Omega^{\mathcal{B}}_0\times[0,T] \vspace{0.15cm}\\
s_2(\bfX,t)=0, & (\bfX,t)\in\partial\Omega^{\mathcal{B}}_0\times[0,T] \vspace{0.15cm}\\
\textbf{s}=\textbf{0}, & \hspace{-1.75cm}(\bfX,t)\in\partial\Omega_0\setminus\left(\partial\Omega^{\mathcal{T}}_0\cup\partial\Omega^{\mathcal{B}}_0\right)\times[0,T]
\end{array}\right. ,\label{BCs}
\end{equation}
where $\bfN$ stands for the outward unit normal to the boundary $\partial\Omega_0$.

\begin{remark}\label{Remark1} The boundary conditions at the grips. {\rm In experiments, ``pure-shear'' specimens are typically gripped in a way that complex triaxial stresses develop near the grips. Numerical experiments indicate that these localized stresses have practically no effect on the response of the specimens, thus our idealized choice of zero traction (\ref{BCs})$_{3,6}$ at the top and bottom boundaries.
}

\end{remark}

\subsubsection{Governing equations}

In the absence of inertia and body forces, putting all the above ingredients together, the mechanical response of the specimens is governed by the equilibrium and incompressibility constraint equations
\begin{equation}
\left\{\begin{array}{ll}{\rm Div}\,\bfS={\bf0}, & \quad (\bfX,t)\in\mathrm{\Omega}_0\times[0,T] \vspace{0.2cm} \\
\det\nabla\bfy=1, & \quad (\bfX,t)\in\mathrm{\Omega}_0\times[0,T]
\end{array}\right. \label{Equilibrium-PDE}
\end{equation}
subject to the initial and boundary conditions (\ref{ICs})$_{1,2}$ and (\ref{BCs}), where $\bfS(\bfX,t)=\mu \nabla\bfy+\nu\nabla\bfy{\bfC^v}^{-1}-p\nabla\bfy^{-T}$, coupled with the evolution equation
\begin{equation}
\dot{\bfC}^v=\dfrac{\nu}{\eta}\left[\nabla\bfy^T\nabla\bfy-
\dfrac{1}{3}\left(\nabla\bfy^T\nabla\bfy\cdot{\bfC^v}^{-1}\right)\bfC^v\right] \label{Evolution-ODE}
\end{equation}
subject to the initial condition (\ref{ICs})$_3$, for the deformation field $\bfy(\bfX,t)$, the pressure field $p(\bfX,t)$, and the internal variable $\bfC^v(\bfX,t)$.

The initial-boundary-value problem (\ref{Equilibrium-PDE})-(\ref{Evolution-ODE}) with (\ref{ICs})-(\ref{BCs}) does not admit analytical solutions and hence must be solved numerically. In a recent contribution, Ghosh et al. \cite{GSKLP21} have introduced a robust scheme based on a finite-element (FE) discretization of space and a high-order finite-difference (FD) discretization of time for such a class of problems. The solutions that we present in the sequel are generated with a variant of that scheme, one where we make use of a non-conforming Crouzeix–Raviart finite-element discretization of first order instead of a conforming one of second order. Also, because of their checked agreement with full 3D solutions, all the solutions that we present in the sequel correspond to plane-stress solutions.

Before proceeding with the presentation of the results, we emphasize that, because of the presence of a pre-existing crack in the specimens, extreme care should be exercised in using a sufficiently refined FE discretization of space and a sufficiently refined FD discretization of time in order to generate converged solutions. All the solutions that are presented below were checked to be converged solutions.

\subsection{Numerical results}

Representative of values for typical elastomers, all the results that follow pertain to equilibrium and non-equilibrium initial shear moduli
\begin{equation*}
\mu=1\; {\rm MPa}\qquad {\rm and}\qquad \nu=2\; {\rm MPa},
\end{equation*}
and three different viscosities:
\begin{equation*}
\eta=5, 10, 20\;  {\rm MPa}\, {\rm s}.
\end{equation*}
Note that these material constants result in elastomers with relaxation times $\tau=\eta/\nu=2.5, 5,$ and $10$ ${\rm s}$. Furthermore, in order to probe the entire spectrum of behaviors --- from elasticity-dominated to viscosity-dominated --- the results correspond to global stretch rates in the range
\begin{equation*}
\dot{\Lambda}_0\equiv\dfrac{\dot{h}_0}{H}\in[10^{-3}, 50]\; {\rm s}^{-1}
\end{equation*}
spanning more than four orders of magnitude.

\subsubsection{The force-deformation response}

\begin{figure}[h!]
  \subfigure[]{
   \label{fig:3a}
   \begin{minipage}[]{0.5\textwidth}
   \centering \includegraphics[width=2.4in]{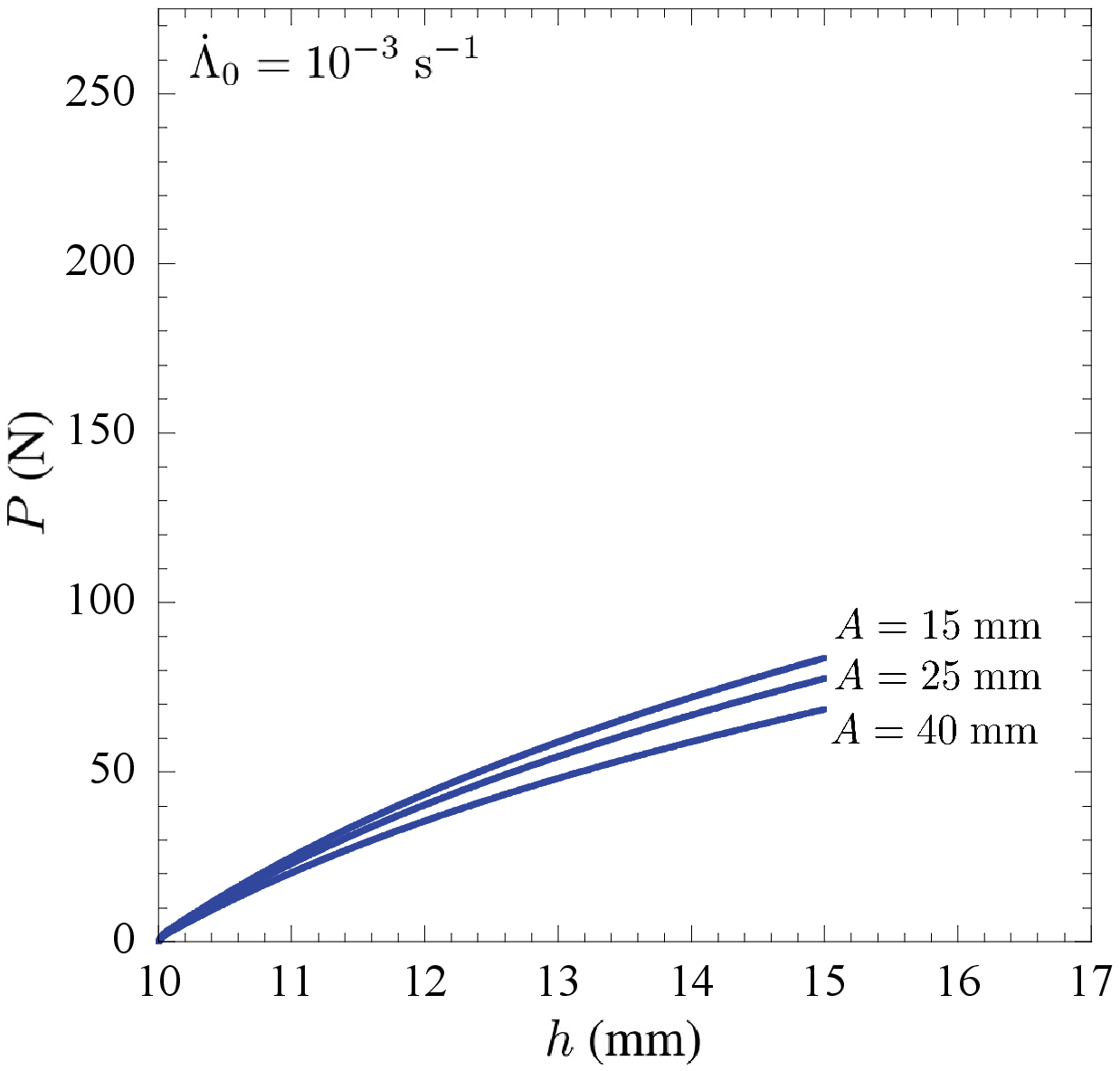}
   \vspace{0.2cm}
   \end{minipage}}
  \subfigure[]{
   \label{fig:3b}
   \begin{minipage}[]{0.5\textwidth}
   \centering \includegraphics[width=2.4in]{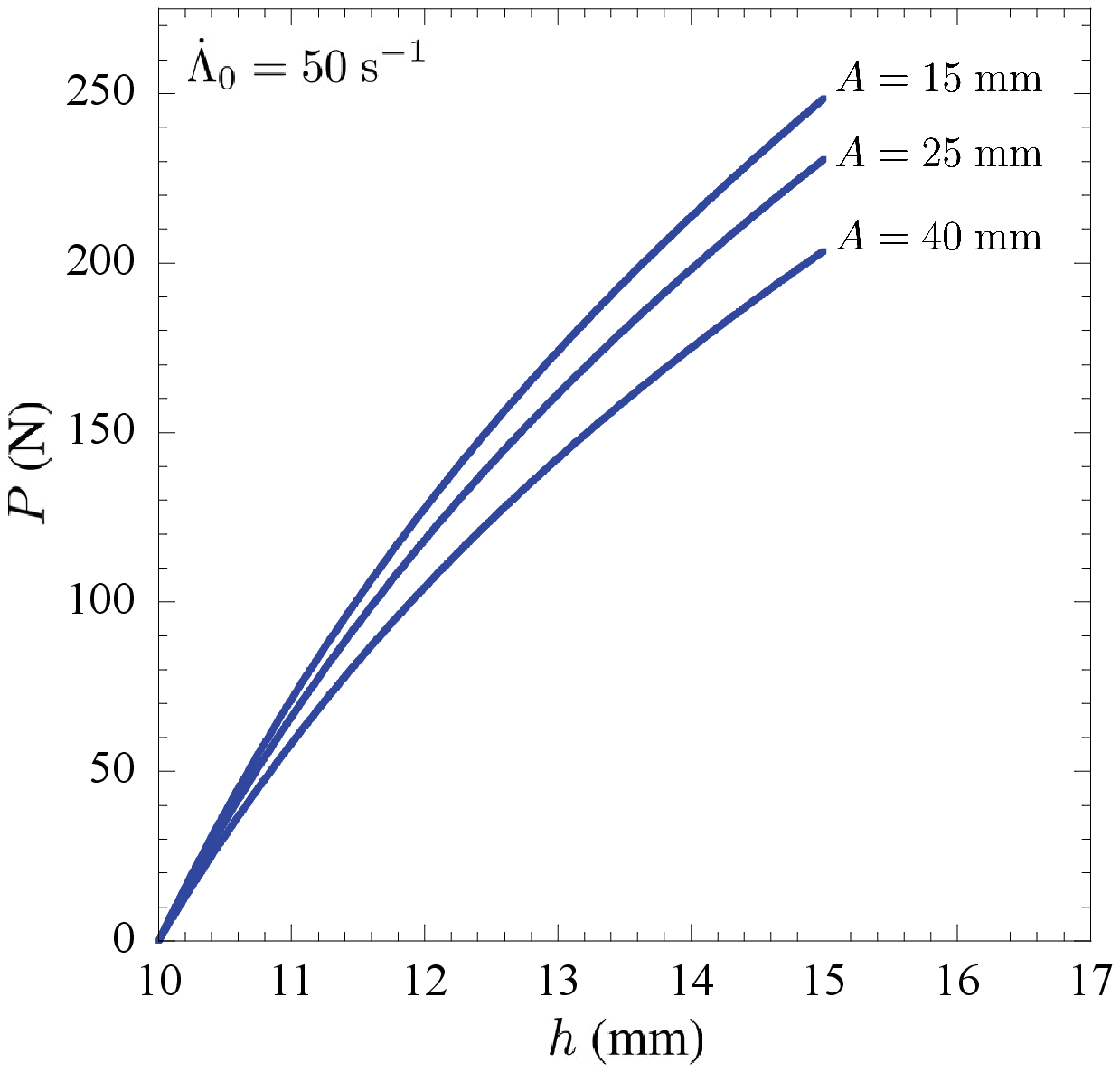}
   \vspace{0.2cm}
   \end{minipage}}
   \caption{Force-deformation response of ``pure-shear'' specimens with viscosity $\eta=5$ ${\rm MPa}\, {\rm s}$ and pre-existing cracks of various lengths $A$. Part (a) shows results for deformations applied at the stretch rate $\dot{\Lambda}_0=10^{-3}$ s$^{-1}$, while part (b) shows results for $\dot{\Lambda}_0=50$ s$^{-1}$.}\label{Fig3}
\end{figure}

Figure \ref{Fig3} presents results for the total force $P$ required to deform the specimens with viscosity $\eta=5$ ${\rm MPa}\, {\rm s}$ and pre-existing cracks of length $A=15, 25, 40$ mm at constant global stretch rates $\dot{\Lambda}_0=10^{-3}$ s$^{-1}$ and $\dot{\Lambda}_0=50$ s$^{-1}$. The results are shown for $P$ as a function of the applied deformation $h$ for $\dot{\Lambda}_0=10^{-3}$ s$^{-1}$ in part (a) and for $\dot{\Lambda}_0=50$ s$^{-1}$ in part (b). Two expected observations are immediate. Specimens with larger cracks require smaller forces to reach the same deformation. Larger forces are required to reach a given deformation applied at a higher global stretch rate.
%
%

\subsubsection{The total deformation energy $\mathcal{W}$ and its partition into $\mathcal{W}^{{\rm Eq}}$, $\mathcal{W}^{{\rm NEq}}$, and $\mathcal{W}^{v}$}

The areas under the curves in the results presented in Fig. \ref{Fig3} correspond to the total work done by the applied loads. By the same token, they correspond to the total deformation energy stored and dissipated by the elastomer. We thus have
\begin{equation*}
\mathcal{W}=\displaystyle\int_{H}^{H(1+\dot{\Lambda}_0 t)} P\,{\rm d}h.
\end{equation*}
%

%
\begin{figure}[b!]
  \subfigure[]{
   \label{fig:4a}
   \begin{minipage}[]{0.5\textwidth}
   \centering \includegraphics[width=2.75in]{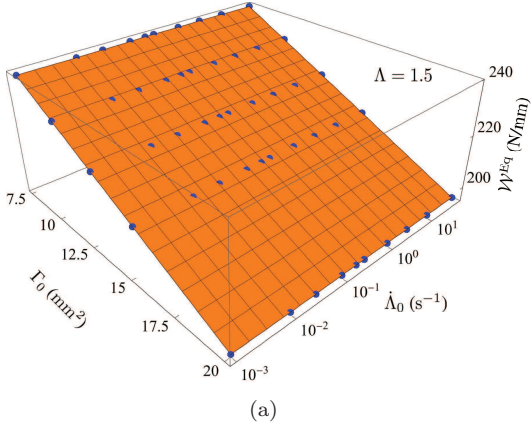}
   \vspace{0.2cm}
   \end{minipage}}
  \subfigure[]{
   \label{fig:4b}
   \begin{minipage}[]{0.5\textwidth}
   \centering \includegraphics[width=2.75in]{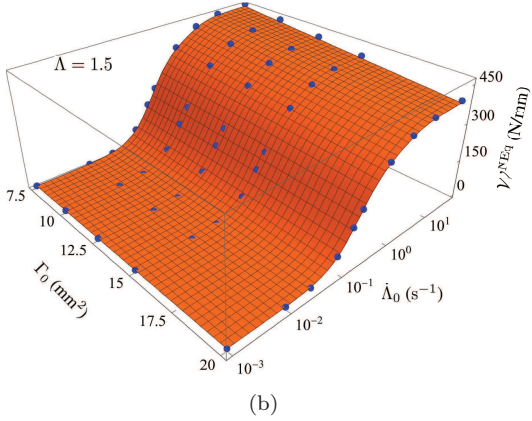}
   \vspace{0.2cm}
   \end{minipage}}
     \subfigure[]{
   \label{fig:4c}
   \begin{minipage}[]{0.5\textwidth}
   \centering \includegraphics[width=2.75in]{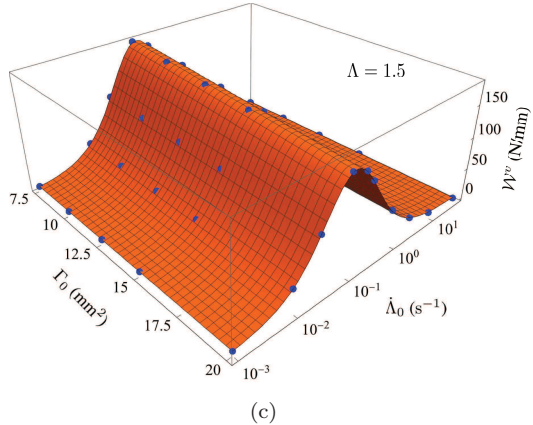}
   \vspace{0.2cm}
   \end{minipage}}
   \caption{Computed values from (\ref{WEq-NH})-(\ref{Wv-NH}) of (a) the equilibrium elastic energy $\mathcal{W}^{{\rm Eq}}$, (b) the non-equilibrium elastic energy $\mathcal{W}^{{\rm NEq}}$, and (c) the dissipated viscous energy $\mathcal{W}^{v}$ in ``pure-shear'' specimens with viscosity $\eta=5$ ${\rm MPa}\, {\rm s}$ stretched at $\Lambda=1.5$, plotted as functions of the initial crack surface $\Gamma_0=A \times B$ and the applied stretch rate $\dot{\Lambda}_0$.}\label{Fig4}
\end{figure}
%
Since the elastomer is a Gaussian elastomer with constant viscosity, we also have that
\begin{align}
\mathcal{W}^{{\rm Eq}}=\displaystyle\int_{\Omega_0}\psi^{{\rm Eq}}(\bfF)\,{\rm d}\bfX=\displaystyle\int_{\Omega_0}\dfrac{\mu}{2}\left[{\rm tr}\,\bfC-3\right]\,{\rm d}\bfX,\label{WEq-NH}
\end{align}
\begin{align}
\mathcal{W}^{{\rm NEq}}=&\displaystyle\int_{\Omega_0}\psi^{{\rm NEq}}(\bfF{\bfF^v}^{-1})\,{\rm d}\bfX\nonumber\\
=&\displaystyle\int_{\Omega_0}\dfrac{\nu}{2}\left[{\rm tr}(\bfC{\bfC^{v}}^{-1})-3\right]\,{\rm d}\bfX, \label{WNEq-NH}
\end{align}
and
\begin{align}
\mathcal{W}^{v}=&\mathcal{W}-\mathcal{W}^{{\rm Eq}}-\mathcal{W}^{{\rm NEq}}\nonumber\\
=&\displaystyle\int_{H}^{H(1+\dot{\Lambda}_0 t)} P\,{\rm d}h-\displaystyle\int_{\Omega_0}\dfrac{\mu}{2}\left[{\rm tr}\,\bfC-3\right]\,{\rm d}\bfX-\nonumber\\
&\displaystyle\int_{\Omega_0}\dfrac{\nu}{2}\left[{\rm tr}(\bfC{\bfC^{v}}^{-1})-3\right]\,{\rm d}\bfX.  \label{Wv-NH}
\end{align}

Figure \ref{Fig4} shows results for $\mathcal{W}^{{\rm Eq}}$, $\mathcal{W}^{{\rm NEq}}$, and $\mathcal{W}^{v}$ --- as computed from expressions (\ref{WEq-NH})-(\ref{Wv-NH}) and the pertinent numerical solutions for the deformation field $\bfy(\bfX,t)$ and internal variable $\bfC^v(\bfX,t)$ --- at the global stretch $\Lambda\equiv h(t)/H=1.5$, plotted as functions of the initial crack surface $\Gamma_0=A \times B$ and the stretch rate $\dot{\Lambda}_0$.

Three comments are in order. First, the results at other fixed values of the stretch $\Lambda$ are not fundamentally different from those shown in Fig. \ref{Fig4} for $\Lambda=1.5$. In other words, the results presented in Fig. \ref{Fig4} can be considered as representative of those at any stretch $\Lambda$. All three parts of the deformation energy appear to be linear with respect to the crack surface $\Gamma_0$. This implies that even the specimen with the smallest pre-existing crack length $A=15$ mm behaves \emph{de facto} as an infinitely long strip containing a semi-infinitely long edge crack. Finally, the dependence of the equilibrium energy $\mathcal{W}^{{\rm Eq}}$ on the applied stretch rate $\dot{\Lambda}_0$ appears to also be linear (as shown in the next subsection, it is in fact constant), while those of the non-equilibrium energy $\mathcal{W}^{{\rm NEq}}$ and the dissipated viscous energy $\mathcal{W}^{v}$ are distinctly nonlinear.

\subsubsection{The derivatives $-\partial\mathcal{W}^{{\rm Eq}}/\partial\Gamma_0$, $-\partial\mathcal{W}^{{\rm NEq}}/\partial\Gamma_0$, and $-\partial\mathcal{W}^{v}/\partial\Gamma_0$}

\begin{figure}[t!]
  \subfigure[]{
   \label{fig:5a}
   \begin{minipage}[]{0.5\textwidth}
   \centering \includegraphics[width=2.4in]{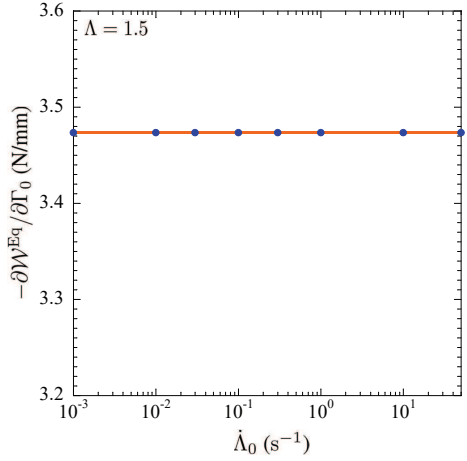}
   \vspace{0.2cm}
   \end{minipage}}
  \subfigure[]{
   \label{fig:5b}
   \begin{minipage}[]{0.5\textwidth}
   \centering \includegraphics[width=2.4in]{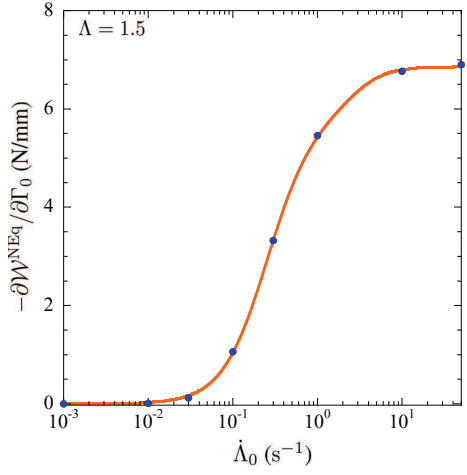}
   \vspace{0.2cm}
   \end{minipage}}
     \subfigure[]{
   \label{fig:5c}
   \begin{minipage}[]{0.5\textwidth}
   \centering \includegraphics[width=2.4in]{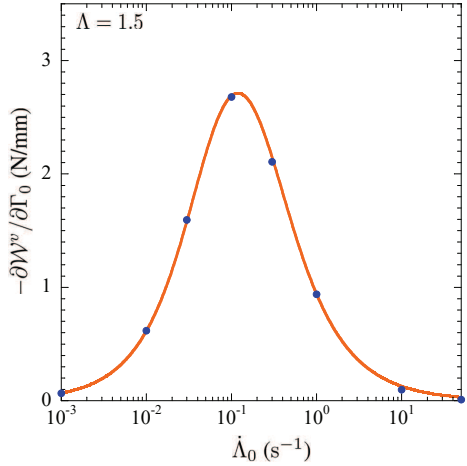}
   \vspace{0.2cm}
   \end{minipage}}
   \caption{Computed values from Fig. \ref{Fig4} of (a) the derivative $-\partial\mathcal{W}^{{\rm Eq}}/\partial\Gamma_0$ of the equilibrium elastic energy, (b) the derivative $-\partial\mathcal{W}^{{\rm NEq}}/\partial\Gamma_0$ of the non-equilibrium elastic energy, and (c) the derivative $-\partial\mathcal{W}^{v}/\partial\Gamma_0$ of the dissipated viscous energy in ``pure-shear'' specimens with viscosity $\eta=5$ ${\rm MPa}\, {\rm s}$ stretched at $\Lambda=1.5$, plotted as functions of the applied stretch rate $\dot{\Lambda}_0$.}\label{Fig5}
\end{figure}

From the 3D plots presented in Fig. \ref{Fig4}, we can readily compute the derivatives entering the general criticality condition (\ref{Tc-1}). The results are presented in Fig. \ref{Fig5} as functions of the applied stretch rate $\dot{\Lambda}_0$. While part (a) presents the results for $-\partial\mathcal{W}^{{\rm Eq}}/\partial\Gamma_0$, parts (b) and (c) present those for $-\partial\mathcal{W}^{{\rm NEq}}/\partial\Gamma_0$ and $-\partial\mathcal{W}^{v}/\partial\Gamma_0$, respectively.

We remark that, consistent with the behavior noted in Fig. \ref{Fig4}, the results in Fig. \ref{Fig5} are invariant with respect to $\Gamma_0$. What is more, the results at other fixed values of the stretch $\Lambda$ are qualitatively the same as those shown in Fig. \ref{Fig5} for $\Lambda=1.5$, which can be therefore viewed as representative of any $\Lambda$.
%
%

We can make two further observations from Fig. \ref{Fig5}. First, consistent with the analysis presented in Section \ref{Sec:Global}, the derivative $-\partial\mathcal{W}^{{\rm Eq}}/\partial\Gamma_0$ is independent of the applied stretch rate $\dot{\Lambda}_0$. Second, and also consistent with the analysis presented in Section \ref{Sec:Global}, the  derivatives $-\partial\mathcal{W}^{{\rm NEq}}/\partial\Gamma_0$ and $-\partial\mathcal{W}^{v}/\partial\Gamma_0$ depend strongly on $\dot{\Lambda}_0$. In particular, as expected on physical grounds, $-\partial\mathcal{W}^{{\rm NEq}}/\partial\Gamma_0$ is bounded from below (by zero) and from above, and increases monotonically with increasing $\dot{\Lambda}_0$. On the other hand, $-\partial\mathcal{W}^{v}/\partial\Gamma_0$ is also bounded from below (by zero) and from above, but is not monotonically increasing in $\dot{\Lambda}_0$, instead, it exhibits a single local maximum at some value of $\dot{\Lambda}_0$ (in the present case, around $\dot{\Lambda}_0=10^{-1}$ ${\rm s}^{-1}$).

The above results for $-\partial\mathcal{W}^{{\rm Eq}}/\partial\Gamma_0$, $-\partial\mathcal{W}^{{\rm NEq}}/\partial\Gamma_0$, and $-\partial\mathcal{W}^{v}/\partial\Gamma_0$, when combined with the experimental fact that viscoelastic elastomers in ``pure-shear'' tests carried out at constant stretch rates nucleate fracture at the same stretch $\Lambda$, corroborate that relations (\ref{G-visc}) and (\ref{Gc-0}) must hold true, for the equality in (\ref{Tc-1}) can be satisfied at fixed $\Lambda$ for all stretch rates $\dot{\Lambda}_0$ only then.

\subsubsection{{\color{black} Scalings of the critical tearing energy $T_c$}}

%
\begin{figure}[b!]
  \subfigure[]{
   \label{fig:6a}
   \begin{minipage}[]{0.5\textwidth}
   \centering \includegraphics[width=2.4in]{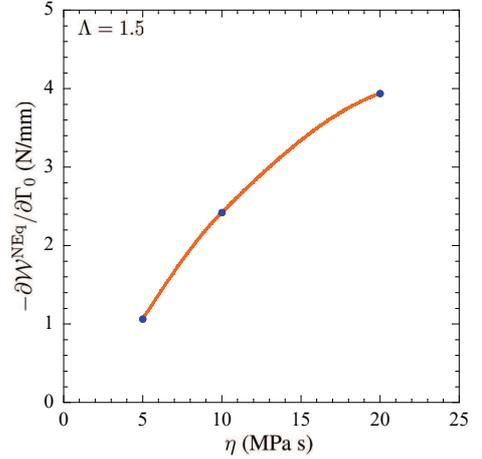}
   \vspace{0.2cm}
   \end{minipage}}
  \subfigure[]{
   \label{fig:6b}
   \begin{minipage}[]{0.5\textwidth}
   \centering \includegraphics[width=2.4in]{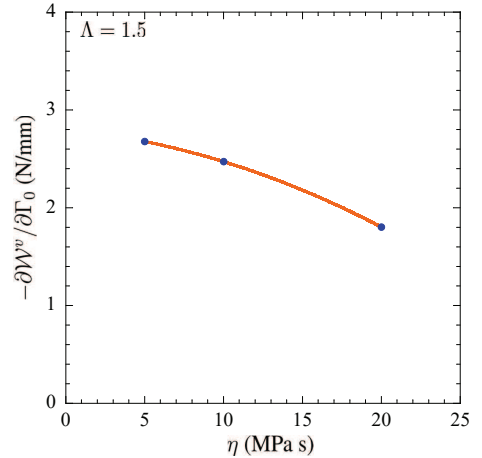}
   \vspace{0.2cm}
   \end{minipage}}
   \caption{Dependence on the viscosity $\eta$ of (a) the derivative $-\partial\mathcal{W}^{{\rm NEq}}/\partial\Gamma_0$ of the non-equilibrium elastic energy and (b) the derivative $-\partial\mathcal{W}^{v}/\partial\Gamma_0$ of the dissipated viscous energy in ``pure-shear'' specimens stretched at $\Lambda=1.5$ at the stretch rate $\dot{\Lambda}_0= 10^{-1}$ ${\rm s}^{-1}$.}\label{Fig6}
\end{figure}
%

{\color{black} While the critical tearing energy $T_c$ has long remained an elusive quantity, it has been known since the 1950s that it scales with the viscosity of the elastomer \cite{Mullins59,Gent96}. It has also been known since the 1970s that $T_c$ scales with the stretch rate in a manner that resembles the dependence of the storage modulus on frequency in DMA (dynamic mechanical analysis) tests of elastomers \cite{Knauss1973,Gent96}.}

Since we have now established that (\ref{G-visc}) holds true and hence --- bringing resolution to the decades-old open problem of how $T_c$ depends on the loading history --- that
\begin{equation*}
T_c=G_c(1+ \,f_c)=G_c-\dfrac{\partial \mathcal{W}^{{\rm NEq}}}{\partial \Gamma_0}-\dfrac{\partial \mathcal{W}^{v}}{\partial \Gamma_0}
\end{equation*}
at fracture, we can readily examine the precise scalings of $T_c$ with the viscosity and the stretch rate for the prototypical elastomer under investigation here.

To reveal the scaling on viscosity, Figs. \ref{Fig6}(a) and \ref{Fig6}(b) plot the values of $-\partial\mathcal{W}^{{\rm NEq}}/\partial\Gamma_0$ and $-\partial\mathcal{W}^{v}/\partial\Gamma_0$ for ``pure-shear'' specimens when stretched at $\Lambda=1.5$ at the stretch rate $\dot{\Lambda}_0= 10^{-1}$ ${\rm s}^{-1}$, as functions of the viscosity $\eta$ of the elastomer, which was varied from $\eta=5$  to $20$ ${\rm MPa}$ ${\rm s}$. Interestingly, for the range of $\eta$ considered, both derivatives scale not far from linearly in $\eta$.

%
\begin{figure}[t!]
   \centering \includegraphics[width=2.4in]{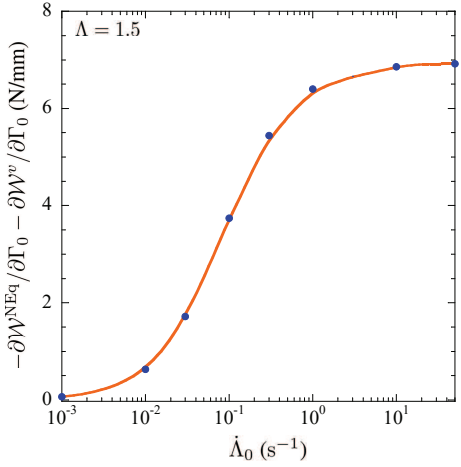}
   \caption{Dependence on the stretch rate $\dot{\Lambda}_0$ of the combination of derivatives $-\partial\mathcal{W}^{{\rm NEq}}/\partial\Gamma_0-\partial\mathcal{W}^{v}/\partial\Gamma_0$ in ``pure-shear'' specimens with viscosity $\eta=5$ ${\rm MPa}\, {\rm s}$ stretched at $\Lambda=1.5$.}\label{Fig7}
\end{figure}
%

{\color{black} To reveal the scaling on stretch rate, Fig. \ref{Fig7} plots the values of the combination of derivatives $-\partial\mathcal{W}^{{\rm NEq}}/\partial\Gamma_0-\partial\mathcal{W}^{v}/\partial\Gamma_0$ for ``pure-shear'' specimens with viscosity $\eta=5$ ${\rm MPa}\, {\rm s}$  when stretched at $\Lambda=1.5$, as a function of the applied stretch rate $\dot{\Lambda}_0$. Consistent with results in the classical literature --- see, for instance, Section 3 in \cite{Gent96} --- the plot does indeed resemble the typical dependence of the storage modulus on frequency obtained from DMA tests of elastomers.}

\subsubsection{The local fields in the regions $\emph{\texttt{A}}$, $\emph{\texttt{B}}$, $\emph{\texttt{C}}$, and $\emph{\texttt{D}}$}

\begin{figure}[t!]
  \subfigure[]{
   \label{fig:7a}
   \begin{minipage}[]{0.45\textwidth}
   \centering \includegraphics[width=3.4in]{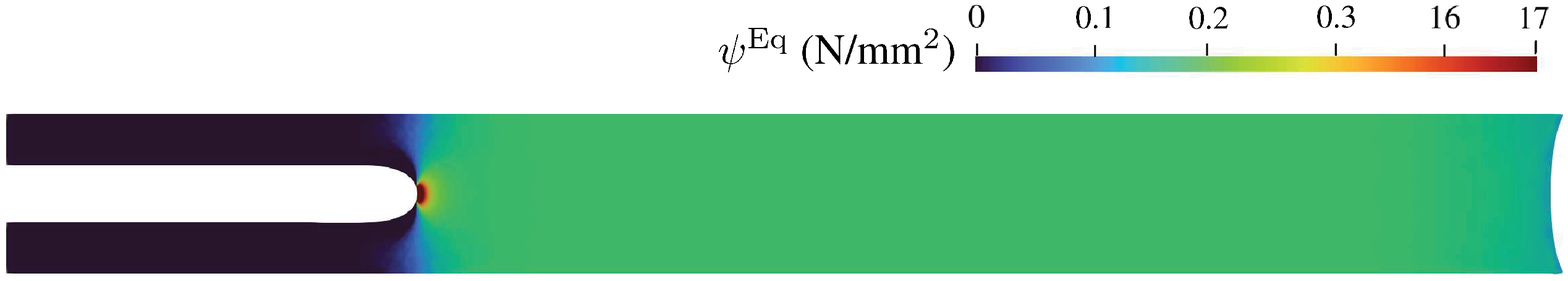}
   \vspace{0.2cm}
   \end{minipage}}
  \subfigure[]{
   \label{fig:7b}
   \begin{minipage}[]{0.45\textwidth}
   \centering \includegraphics[width=3.4in]{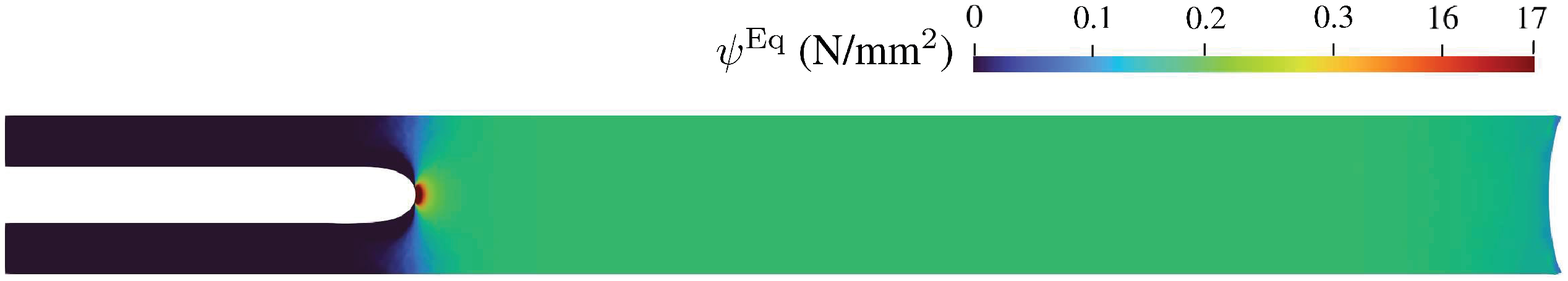}
   \vspace{0.2cm}
   \end{minipage}}
   \caption{Contour plots over the deformed configuration of the equilibrium elastic energy density $\psi^{{\rm Eq}}(\bfF)$ in ``pure-shear'' specimens with viscosity $\eta=5$ ${\rm MPa}$ ${\rm s}$ and pre-existing crack of length $A=40$ mm stretched at $\Lambda=1.5$ at stretch rates (a) $\dot{\Lambda}_0= 10^{-3}$ ${\rm s}^{-1}$ and (b) $\dot{\Lambda}_0= 50$ ${\rm s}^{-1}$.}\label{Fig8}
\end{figure}

For completeness, we close this section by reporting in Fig. \ref{Fig8} representative contour plots of the equilibrium elastic energy density $\psi^{{\rm Eq}}(\bfF)$ in specimens stretched at $\Lambda=1.5$ at two different stretch rates, $\dot{\Lambda}_0= 10^{-3}$ ${\rm s}^{-1}$ and $\dot{\Lambda}_0= 50$ ${\rm s}^{-1}$. The results pertain to an elastomer with viscosity $\eta=5$ ${\rm MPa}$ ${\rm s}$, a pre-existing crack of length $A=40$ mm, and are shown over the deformed configuration.

These plots allow to identify the precise locations of the so-called regions $\texttt{A}$, $\texttt{B}$, $\texttt{C}$, and $\texttt{D}$ in the global Rivlin-Thomas analysis of the problem. They provide as well quantitative insight into the spatial heterogeneity of the local deformation field in the crack-front region $\texttt{B}$ and the fringe region $\texttt{D}$. Consistent with the results in Fig. \ref{Fig4}(a) for the total equilibrium elastic energy $\mathcal{W}^{{\rm Eq}}$, note that the local value of $\psi^{{\rm Eq}}(\bfF)$ is independent of the applied stretch rate $\dot{\Lambda}_0$ over the entire specimen.

%
%

{\color{black}

\section{Comparisons with experiments on VHB 4905}\label{Sec:Experiments}

In this section, as a first demonstration of its use to explain fracture in elastomers, we deploy the Griffith criticality condition (\ref{Gc-0}) to explain a representative set of ``pure-shear'' fracture tests, those reported in \cite{Pharretal2012} on the acrylic elastomer VHB 4905. The focus is on the results for specimens with the same geometry considered in the preceding section ($L=152$ mm, $H=10$ mm, $B=0.5$ mm) featuring a pre-existing edge crack of length $A=20$ mm; see Fig. 3(b) in \cite{Pharretal2012}.

\subsection{The viscoelastic behavior of VHB 4905}

%
\begin{figure}[t!]
  \subfigure[]{
   \label{fig:8a}
   \begin{minipage}[]{0.5\textwidth}
   \centering \includegraphics[width=2.4in]{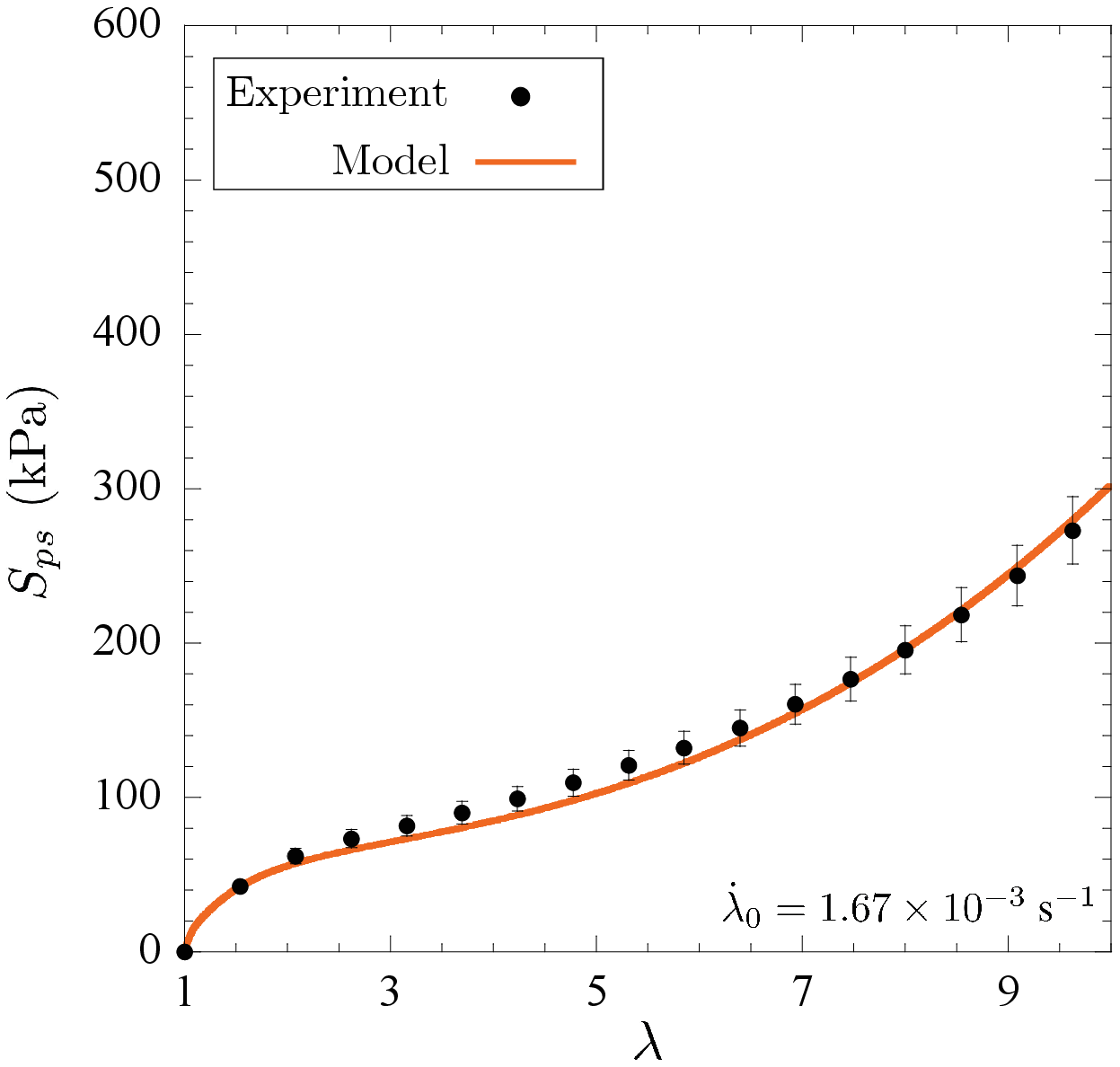}
   \vspace{0.2cm}
   \end{minipage}}
  \subfigure[]{
   \label{fig:8b}
   \begin{minipage}[]{0.5\textwidth}
   \centering \includegraphics[width=2.4in]{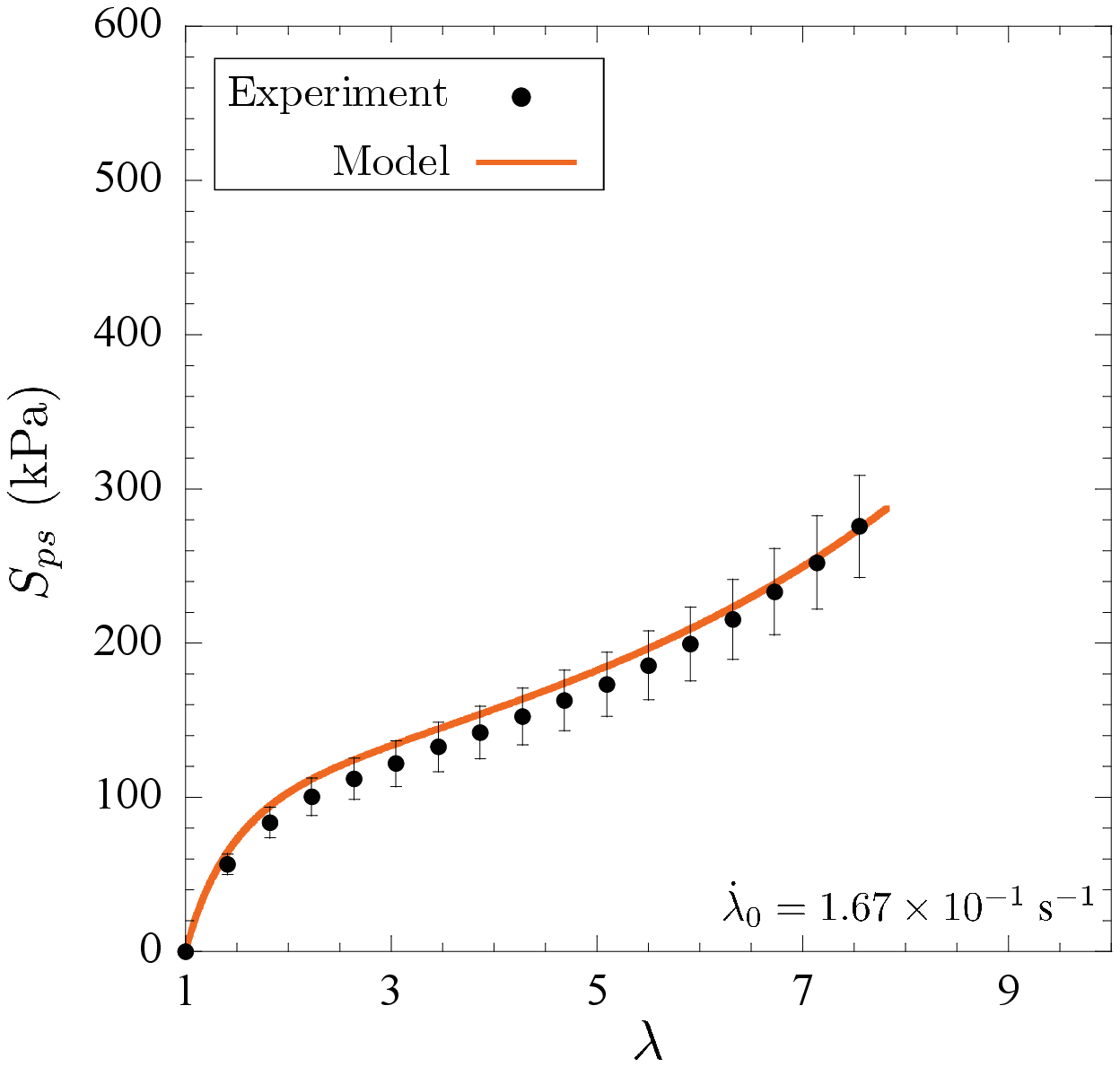}
   \vspace{0.2cm}
   \end{minipage}}
     \subfigure[]{
   \label{fig:8c}
   \begin{minipage}[]{0.5\textwidth}
   \centering \includegraphics[width=2.4in]{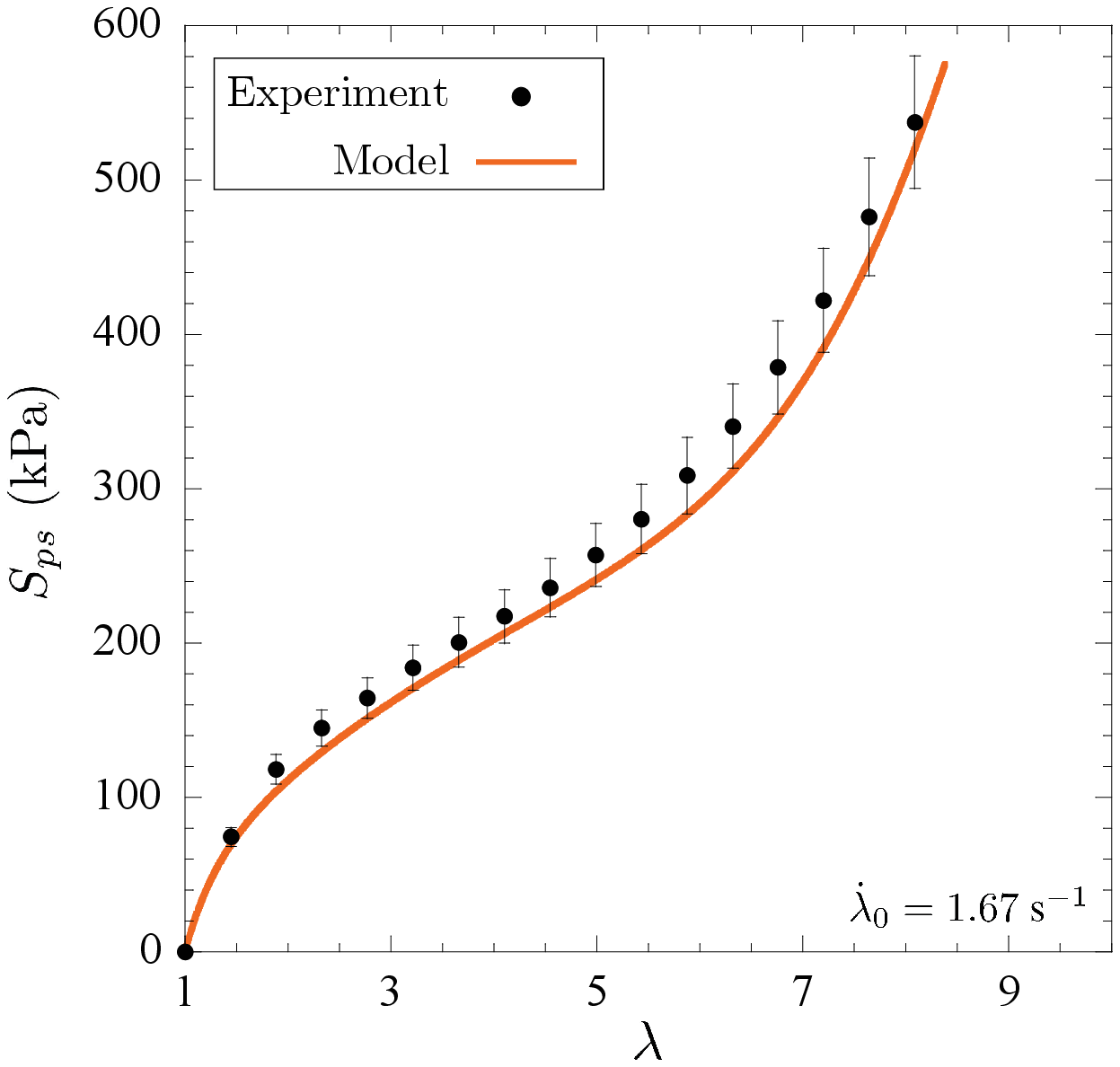}
   \vspace{0.2cm}
   \end{minipage}}
   \caption{Comparison between the stress-stretch response (solid line) predicted by the viscoelastic model (\ref{S-KLP})-(\ref{Evolution-KLP}), with the material constants in Table \ref{Table1}, and the experimental data (solids circles) reported in \cite{Pharretal2012} for the acrylic elastomer VHB 4905 under pure shear applied at three different constant stretch rates: (a) $\dot{\lambda}_0= 1.67\times 10^{-3}$ ${\rm s}^{-1}$, (b) $\dot{\lambda}_0= 1.67\times 10^{-1}$ ${\rm s}^{-1}$, and (c) $\dot{\lambda}_0= 1.67$ ${\rm s}^{-1}$.}\label{Fig9}
\end{figure}
%
In contrast to the canonical viscoelastic behavior considered above, VHB 4905 exhibits non-Gaussian elasticity and a nonlinear viscosity of shear-thinning type. This falls squarely within the behavior of the vast majority of elastomers \cite{Treloar75,Doi98,Gent62,KOLP02,Lion06,LP10,GLP2021,Cohen21}. Such a behavior can be described with the same type of two-potential constitutive modelling considered in the preceding section by simply replacing the finite branch of the equilibrium and non-equilibrium Gaussian free-energy functions in (\ref{psi}) with the non-Gaussian free-energy functions
\begin{equation*}
\left\{\hspace{-0.1cm}\begin{array}{l}
\psi^{{\rm Eq}}(\bfF)=\displaystyle\sum\limits_{r=1}^2\dfrac{3^{1-\alpha_r}}{2 \alpha_r}\mu_r\left[\,I_1^{\alpha_r}-3^{\alpha_r}\right]\vspace{0.2cm}\\
\psi^{{\rm NEq}}(\bfF{\bfF^{v}}^{-1})=\displaystyle\sum\limits_{r=1}^2\dfrac{3^{1-\beta_r}}{2 \beta_r}\nu_r\left[\,{I^e_1}^{\beta_r}-3^{\beta_r}\right]
\end{array}\right. , 
\end{equation*}
and by replacing the constant viscosity in (\ref{phi}) with the nonlinear viscosity function
\begin{equation*}
\eta(I_1^e,I_2^e,I_1^v)=\eta_{\infty}+\dfrac{\eta_0-\eta_{\infty}+K_1\left[{I_1^v}^{\gamma_1}-3^{\gamma_1}\right]}{1+\left(K_2 \mathcal{J}_2^{{\rm NEq}}\right)^{\gamma_2}}.
\end{equation*}
In these expressions, $I^v_1={\rm tr}\,\bfC^v$, $\mathcal{J}_2^{{\rm NEq}}=({I_1^e}^2/3-I_2^e)\times$ $(\sum_{r=1}^2 3^{1-\beta_r}\nu_r {I_1^e}^{\beta_r-1})^2$,  $I_2^e=\frac{1}{2}[(\bfC\cdot{\bfC^v}^{-1})^2-{\bfC^v}^{-1}\bfC\cdot\bfC{\bfC^v}^{-1}]$, and we recall that $I_1$ and $I^e_1$ stand for the invariants (\ref{invariants})$_{1,3}$. Making use of these constitutive prescriptions in (\ref{S-gen})-(\ref{Evolution-gen}) results in the viscoelastic model \cite{KLP16}

\begin{align}
\bfS(\bfX,t)=&\displaystyle\sum\limits_{r=1}^2 3^{1-\alpha_r} \mu_r I_1^{\alpha_r-1} \bfF+\nonumber\\
&\displaystyle\sum\limits_{r=1}^2 3^{1-\beta_r}\nu_r {I^e_1}^{\beta_r-1}\bfF{\bfC^v}^{-1}-p\bfF^{-T},\label{S-KLP}
\end{align}
where, again, $p$ stands for the arbitrary hydrostatic pressure associated with the assumed incompressibility of the elastomer and where now the internal variable $\bfC^v$ is defined implicitly as the solution of the evolution equation
\begin{align}
\dot{\bfC}^v(\bfX,t)=&\dfrac{\sum\limits_{r=1}^2 3^{1-\beta_r}\nu_r {I^e_1}^{\beta_r-1}}{\eta(I_1^e,I_2^e,I_1^v)}\left[\bfC-\dfrac{1}{3}\left(\bfC\cdot{\bfC^v}^{-1}\right)\bfC^v\right]. \label{Evolution-KLP}
\end{align}

In all, the constitutive relation (\ref{S-KLP})-(\ref{Evolution-KLP}) contains fourteen material constants. Four of them, $\mu_1$, $\mu_2$, $\alpha_1$, $\alpha_2$, serve to characterize the non-Gaussian elasticity of the elastomer at states of thermodynamic equilibrium. Another four, $\nu_1$, $\nu_2$, $\beta_1$, $\beta_2$, characterize the non-Gaussian elasticity  at non-equilibrium states. The last six constants, $\eta_0$, $\eta_{\infty}$, $K_1$, $K_2$, $\gamma_1$, $\gamma_2$, serve to characterize the nonlinear shear-thinning viscosity. Note that the constitutive relation (\ref{S-KLP})-(\ref{Evolution-KLP}) includes ($\alpha_1=\beta_1=1$, $\mu_2=\nu_2=0$, $\eta_{\infty}=0$, $K_1=K_2=0$)  the canonical relation (\ref{S-Neo})-(\ref{Evolution-Neo}) for a viscoelastic elastomer with Gaussian elasticity and constant viscosity as a special case.

%
%

In addition to reporting results for ``pure-shear'' fracture tests, the work \cite{Pharretal2012} includes experimental results for the stress-stretch response of VHB 4905 under (approximately) pure shear deformation applied at various constant stretch rates in the range $\dot{\lambda}_0\in[1.67\times 10^{-3},1.67]$ ${\rm s}^{-1}$; see Fig. 3(a) in \cite{Pharretal2012}. Specializing the constitutive relation (\ref{S-KLP})-(\ref{Evolution-KLP}) to such loadings --- that is, to deformation gradients of the form $\bfF={\rm diag}(\lambda,\lambda^{-1},1)$ with $\lambda=1+\dot{\lambda}_{0} t$ and first Piola-Kirchhoff stresses of the form $\bfS={\rm diag}(S_{ps},0,S_{lat})$ --- and then fitting (by least squares) its material constants to the experimental data in \cite{Pharretal2012} yields the values listed in Table \ref{Table1} for all fourteen material constants. As seen from the comparisons presented in Fig. \ref{Fig9}, the constitutive relation (\ref{S-KLP})-(\ref{Evolution-KLP}) with such material constants describes reasonably well the viscoelastic response of VHB 4905 reported in \cite{Pharretal2012}.
\begin{table}[h!]\centering
{\color{black}
\caption{{\color{black} Values of the material constants in the viscoelastic model (\ref{S-KLP})-(\ref{Evolution-KLP}) for the acrylic elastomer VHB 4905.}}
\begin{tabular}{llll}
\hline
$\mu_1=13.96$ kPa & & & $\mu_2=0.9255$ kPa  \\
$\alpha_1=0.5104$ & & & $\alpha_2=1.910$  \\
$\nu_1=50.15$ kPa & & & $\nu_2=5.193\times10^{-6}$ kPa \\
$\beta_1=0.9660$ & & & $\beta_2=7.107$ \\
$\eta_0=7007$ kPa s & & & $\eta_{\infty}=14$ kPa s \\
$K_1=2833$ kPa s & & & $K_2=1.228$ kPa$^{-2}$  \\
$\gamma_1=3.467$ & & & $\gamma_2=0.0836$ \\
\hline
\end{tabular} \label{Table1} }
\end{table}

\subsection{Computation of the derivative $-\partial\mathcal{W}^{{\rm Eq}}/\partial\Gamma_0$}

%
\begin{figure}[b!]
\centering \includegraphics[width=2.4in]{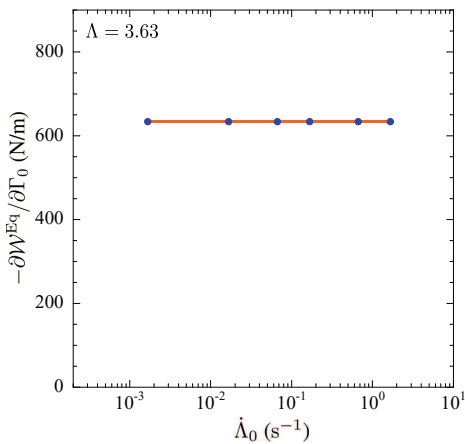}
\caption{The derivative $-\partial\mathcal{W}^{{\rm Eq}}/\partial\Gamma_0$ of the equilibrium elastic energy computed from the simulations of ``pure-shear'' fracture tests on VHB 4905. The results correspond to a global stretch of $\Lambda=3.63$ and are plotted as a function of the applied stretch rate $\dot{\Lambda}_0$.}\label{Fig10}
\end{figure}
%
Having determined the viscoelastic behavior of VHB 4905, we proceed by repeating the same type of full-field analysis presented in Section \ref{Sec:Full-field} in order to compute the derivative $-\partial\mathcal{W}^{{\rm Eq}}/\partial\Gamma_0$ entering the Griffith criticality condition (\ref{Gc-0}).

Before presenting and discussing the results for $-\partial\mathcal{W}^{{\rm Eq}}/$ $\partial\Gamma_0$, a few technical remarks are in order. Since the experiments in \cite{Pharretal2012} pertain to specimens with a pre-existing edge crack of length $A=20$ mm, we perform the simulations for specimens with three crack lengths, $A=15, 20, 25$ mm. This suffices to be able to take the required derivative $-\partial\mathcal{W}^{{\rm Eq}}/\partial\Gamma_0$ in (\ref{Gc-0}) at $\Gamma_0=A\times B=20\times 0.5$ mm$^2$. Much like the global stretch rates used in the experiments, we carry out simulations at six different global stretch rates, $\dot{\Lambda}_0=1.67\times 10^{-3},  1.67\times 10^{-2}, 6.67\times 10^{-2}, 1.67\times 10^{-1}, 6.67\times 10^{-1}, 1.67$ s$^{-1}$. Accordingly, in all, we carry out $3\times 6=18$ simulations of the ``pure-shear'' fracture tests. Furthermore, since the experiments indicate that fracture nucleates from the pre-existing crack at the critical global stretch $\Lambda_c\equiv h_c/H=3.63\pm0.45$, we carry out each of these simulations up to a global stretch of $\Lambda=3.63$.

Analogous to Fig. \ref{Fig5}(a), Fig. \ref{Fig10} presents results for the derivative $-\partial\mathcal{W}^{{\rm Eq}}/\partial\Gamma_0$ computed from the simulations of the ``pure-shear'' fracture tests on VHB 4905, at the global applied stretch $\Lambda=3.63$. Much like the results in Fig. \ref{Fig5}(a) for the canonical case of an elastomer with Gaussian elasticity and constant viscosity, the results in Fig. \ref{Fig10} are invariant with respect to $\Gamma_0$ and independent of the applied stretch rate $\dot{\Lambda}_0$. According to the Griffith criticality condition (\ref{Gc-0}), they indicate then that fracture nucleates at $\Lambda=3.63$ for all $\dot{\Lambda}_0$ precisely as in the experiments, so long as the intrinsic fracture energy of VHB 4905 is about $G_c=634$ N/m.

\begin{remark}\label{Remark3} The value of $G_c=634$ \emph{N/m} for VHB 4905. {\rm The value $G_c=634$ N/m for the intrinsic fracture energy deduced from Fig. \ref{Fig10} depends directly on the constitutive relation (\ref{S-KLP})-(\ref{Evolution-KLP}), together with the material constants in Table \ref{Table1}, utilized in the simulations to describe the viscoelastic behavior of VHB 4905. More specifically, it depends on the part of the model that describes the equilibrium elasticity and hence, here, on the material constants $\mu_1$, $\mu_2$, $\alpha_1$, $\alpha_2$. It is possible that fitting a set of experimental results larger than the one fitted here could lead to material constants $\mu_1$, $\mu_2$, $\alpha_1$, $\alpha_2$ different from those listed in Table \ref{Table1} that describe more accurately the equilibrium elasticity of VHB 4905.

Be that as it may, the comparisons presented in Fig. \ref{Fig9} suggest that the constitutive relation (\ref{S-KLP})-(\ref{Evolution-KLP}), with the material constants in Table \ref{Table1}, describes fairly accurately the viscoelastic behavior of VHB 4905 and hence that the value $G_c=634$ N/m obtained in this work should be a good estimate. Interestingly, this value is considerably larger than those found for common hydrocarbon elastomers, which, again, typically fall within the range (\ref{Gc-range}). This result, we hope, will encourage experiments in the spirit of those carried out in \cite{Ahagon-Gent75,Gent82} to measure directly the value of $G_c$ for VHB 4905.
}

\end{remark}

\begin{remark}\label{Remark4} A Rivlin-Thomas-type formula. {\rm For a ``pure-shear'' fracture test, the computation of the derivative $-\partial\mathcal{W}^{{\rm Eq}}/\partial\Gamma_0$ in the Griffith criticality condition (\ref{Gc-0}) requires, in principle, the numerical solution of the pertinent boundary-value problem. This is precisely the approach that we have followed in this and in the preceding section.

Fortunately, as already alluded to at the end of Section \ref{Sec:Global} above, the approximate formula originally worked out by Rivlin and Thomas \cite{Rivlin1953} in the setting of finite elasticity also happens to apply, \emph{mutatis mutandis}, in the present setting of finite viscoelasticity to compute $-\partial\mathcal{W}^{{\rm Eq}}/\partial\Gamma_0$. Precisely, when the viscoelasticity of the elastomer of interest is described within the two-potential framework, the formula reads
\begin{align*}
-\dfrac{\partial\mathcal{W}^{{\rm Eq}}}{\partial\Gamma_0}=H\psi^{{\rm Eq}}(\bfF_{ps}),\quad \bfF_{ps}={\rm diag}(\lambda,\lambda^{-1},1),
\end{align*}
where we recall that $H$ denotes the initial height of the specimen; see Fig. \ref{Fig1}. This approximate relation --- which we have checked to be in good agreement with all the numerical results that we have generated for $-\partial\mathcal{W}^{{\rm Eq}}/\partial\Gamma_0$ in this Letter --- is obviously of great practical utility as it allows to determine the energy release rate  $-\partial\mathcal{W}^{{\rm Eq}}/\partial\Gamma_0$ in ``pure-shear'' fracture tests solely from knowledge of the equilibrium elasticity of the elastomer and the initial geometry of the specimen without having to solve any boundary-value problem.
}

\end{remark}

\subsection{The critical stretch and the critical stress at fracture: Theory vs. experiment}

At this stage, we are ready to deploy the Griffith criticality condition (\ref{Gc-0}) to explain the ``pure-shear'' fracture tests in \cite{Pharretal2012}. Taking $G_c=634$ N/m as the intrinsic fracture energy of VHB 4905, Fig. \ref{Fig11} confronts the predictions obtained from the simulations for the critical global stretch $\Lambda_c=h_c/H$ in part (a) and the critical global stress $S_c=P_c/BL$ in part (b) at which fracture nucleates, according to the Griffith criticality condition (\ref{Gc-0}), with the corresponding experimental results reported in \cite{Pharretal2012}. The results are presented as functions of the global stretch rate $\dot{\Lambda}_0$ at which the tests are carried out.
%
\begin{figure}[h!]
  \subfigure[]{
   \label{fig:10a}
   \begin{minipage}[]{0.5\textwidth}
   \centering \includegraphics[width=2.4in]{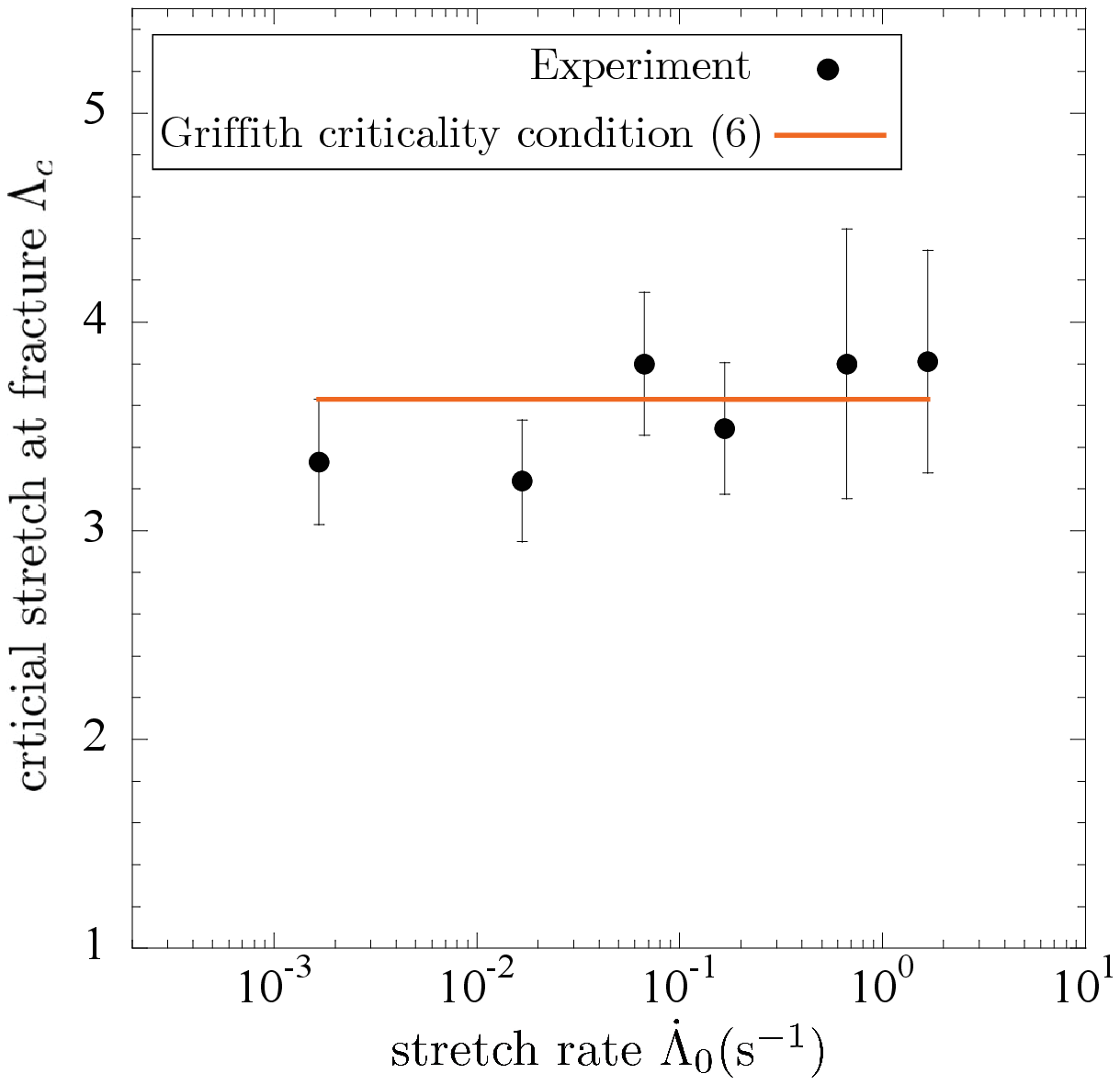}
   \vspace{0.2cm}
   \end{minipage}}
  \subfigure[]{
   \label{fig:10b}
   \begin{minipage}[]{0.5\textwidth}
   \centering \includegraphics[width=2.4in]{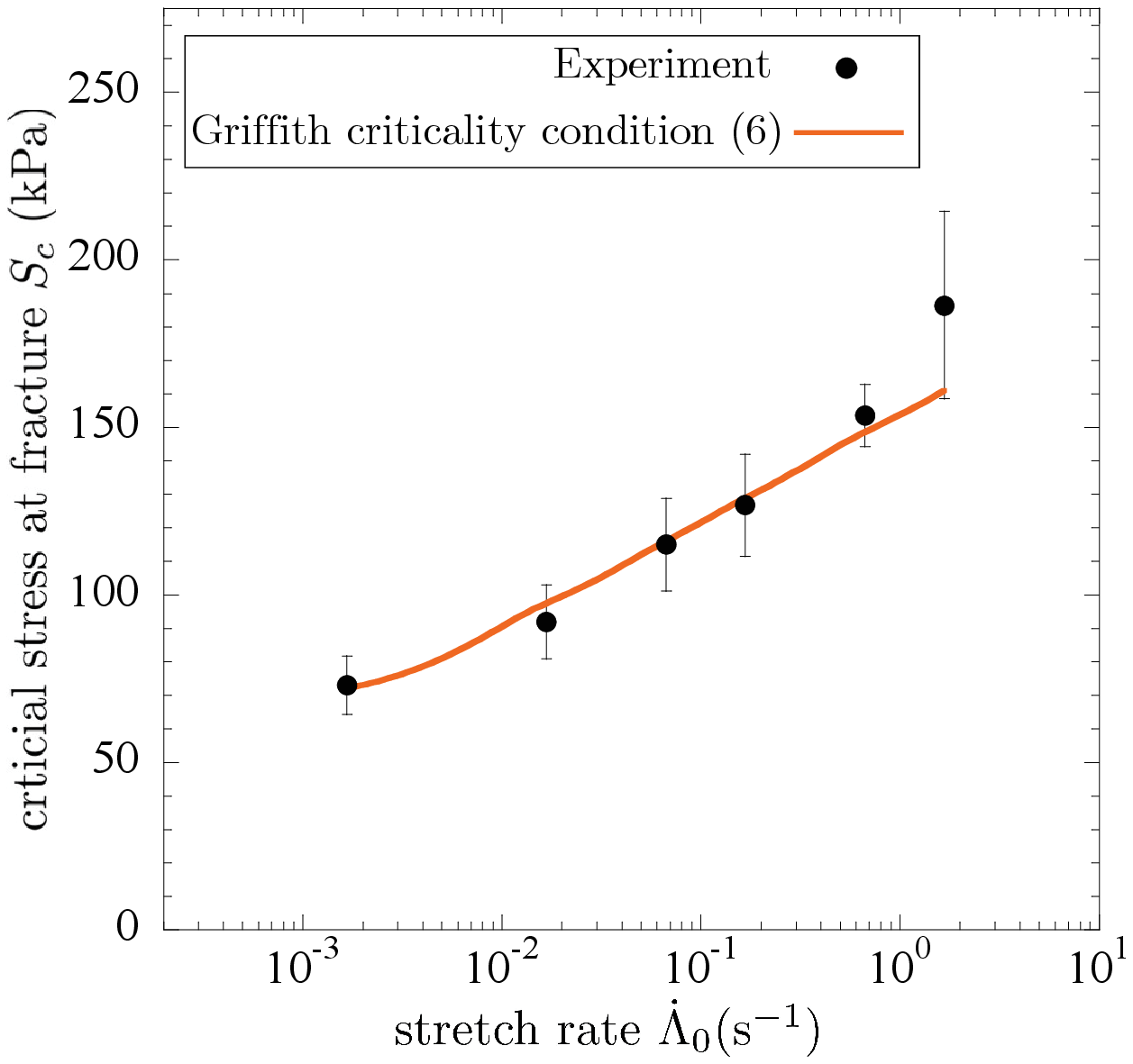}
   \vspace{0.2cm}
   \end{minipage}}
   \caption{Comparison between (a) the critical global stretch $\Lambda_c=h_c/H$ and (b) the critical global stress $S_c=P_c/B L$ at which fracture nucleates, according to the Griffith criticality condition (\ref{Gc-0}), and the corresponding experimental results reported in \cite{Pharretal2012} for ``pure-shear'' fracture tests on VHB  4905. Both sets of results are presented as functions of the global stretch rate $\dot{\Lambda}_0=\dot{h}_0/H$ at which the tests are carried out.}\label{Fig11}
\end{figure}
%

Two observations are immediate from Fig. \ref{Fig11}. First are foremost, the Griffith criticality condition (\ref{Gc-0}) does indeed determine when fracture nucleates from the pre-existing crack in the specimens. Second, as opposed to the critical global stretch $\Lambda_c$, the critical global stress $S_c$ is strongly dependent on the stretch rate $\dot{\Lambda}_0$ at which the tests are carried out, in particular, $S_c$ increases with increasing $\dot{\Lambda}_0$. This dependence is nothing more than a manifestation of the viscoelastic behavior of VHB  4905.

}

\section{Final comments}\label{Sec:Final_Comments}

Besides the ``pure-shear'' fracture test examined in this work, there is another classical test in the literature that provides an additional elementary experimental check of the validity of (\ref{Gc-0}) as the true Griffith condition for viscoelastic elastomers, that is the delayed fracture test of an elastomer sheet, containing a crack, subjected to a constant load that is applied rapidly and then kept fixed. Indeed, the existence of a time delay after the application of the load for the nucleation of fracture from the pre-existing crack in these tests is a telltale of the validity of (\ref{Gc-0}). This is because a time delay implies that it is the increase of $\mathcal{W}^{{\rm Eq}}$ in time at the expense of the decrease of $\mathcal{W}^{{\rm NEq}}$ due to the creeping of the elastomer that leads to the attainability of the criticality condition (\ref{Gc-0}). The first experiments that showed that elastomers exhibit delayed crack growth can be traced back to work of Knauss \cite{Knauss70} in the 1970s. In a companion paper \cite{SLP22}, we explain the pioneering delayed fracture tests of Knauss \cite{Knauss70} and in so doing describe the use of the fundamental form (\ref{Gc-0}) of the Griffith criticality condition when the applied boundary conditions are traction boundary conditions.

We conclude by {\color{black} emphasizing that, from a practical point of view, as illustrated in Section \ref{Sec:Experiments}, the Griffith criticality condition (\ref{Gc-0}) is straightforward to employ, as it is based on quantities that can be measured experimentally once and for all by means of conventional tests.} From a theoretical point of view, moreover, it would behoove us to investigate whether the alluringly simple and intuitive form (\ref{Gc-0}) remains applicable to dissipative solids at large, not just viscoelastic elastomers.

\section*{Acknowledgements}

This work was supported by the National Science Foundation through the Grant CMMI--1901583. This support is gratefully acknowledged.

\end{document}